\PassOptionsToPackage{unicode}{hyperref}
\PassOptionsToPackage{hyphens}{url}
\PassOptionsToPackage{dvipsnames,svgnames,x11names}{xcolor}
\documentclass[
  10pt,
]{article}
\usepackage{fontspec} 
\usepackage{xcolor}
\usepackage{amsmath,amssymb}
\usepackage{iftex}
\ifPDFTeX
  \usepackage[T1]{fontenc}
  \usepackage[utf8]{inputenc}
  \usepackage{textcomp} 
\else 
  \usepackage{unicode-math} 
  \defaultfontfeatures{Scale=MatchLowercase}
  \defaultfontfeatures[\rmfamily]{Ligatures=TeX,Scale=1}
\fi
\usepackage{lmodern}
\ifPDFTeX\else
\fi
\IfFileExists{upquote.sty}{\usepackage{upquote}}{}
\IfFileExists{microtype.sty}{
  \usepackage[]{microtype}
  \UseMicrotypeSet[protrusion]{basicmath} 
}{}
\makeatletter
\@ifundefined{KOMAClassName}{
  \IfFileExists{parskip.sty}{%
    \usepackage{parskip}
  }{
    \setlength{\parindent}{0pt}
    \setlength{\parskip}{6pt plus 2pt minus 1pt}}
}{
  \KOMAoptions{parskip=half}}
\makeatother
\usepackage{longtable,booktabs,array}
\usepackage{calc} 
\usepackage{etoolbox}
\makeatletter
\patchcmd\longtable{\par}{\if@noskipsec\mbox{}\fi\par}{}{}
\makeatother
\IfFileExists{footnotehyper.sty}{\usepackage{footnotehyper}}{\usepackage{footnote}}
\makesavenoteenv{longtable}
\usepackage{graphicx}
\makeatletter
\newsavebox\pandoc@box
\newcommand*\pandocbounded[1]{
  \sbox\pandoc@box{#1}%
  \Gscale@div\@tempa{\textheight}{\dimexpr\ht\pandoc@box+\dp\pandoc@box\relax}%
  \Gscale@div\@tempb{\linewidth}{\wd\pandoc@box}%
  \ifdim\@tempb\p@<\@tempa\p@\let\@tempa\@tempb\fi
  \ifdim\@tempa\p@<\p@\scalebox{\@tempa}{\usebox\pandoc@box}%
  \else\usebox{\pandoc@box}%
  \fi%
}
\def\fps@figure{htbp}
\makeatother
\NewDocumentCommand\citeproctext{}{}

\makeatletter
 \let\@cite@ofmt\@firstofone
 \def\@biblabel#1{}
 \def\@cite#1#2{{#1\if@tempswa , #2\fi}}
\makeatother
\newlength{\cslhangindent}
\newlength{\csllabelwidth}
\newenvironment{CSLReferences}[2] 
 {\begin{list}{}{%
  \setlength{\itemindent}{0pt}
  \setlength{\leftmargin}{0pt}
  \setlength{\parsep}{0pt}
  \ifodd #1
   \setlength{\leftmargin}{\cslhangindent}
   \setlength{\itemindent}{-1\cslhangindent}
  \fi
  \setlength{\itemsep}{#2\baselineskip}}}
 {\end{list}}
\usepackage{calc}

\usepackage[letterpaper,top=1in,bottom=1in,left=1.25in,right=1.25in]{geometry}
\usepackage{microtype}
\usepackage{booktabs}
\usepackage{array}
\usepackage{graphicx}
\setkeys{Gin}{width=0.82\linewidth}
\usepackage{caption}
\usepackage{xcolor}
\usepackage{titlesec}
\titleformat{\section}{\large\bfseries}{\thesection}{0.7em}{}
\titlespacing*{\section}{0pt}{1.9ex plus .3ex}{0.8ex}
\titleformat{\subsection}{\normalsize\bfseries}{\thesubsection}{0.6em}{}
\titlespacing*{\subsection}{0pt}{1.2ex plus .2ex}{0.4ex}
\titleformat{\subsubsection}{\normalsize\bfseries\itshape}{\thesubsubsection}{0.5em}{}
\titlespacing*{\subsubsection}{0pt}{0.9ex}{0.3ex}
\usepackage{fancyhdr}
\fancypagestyle{plain}{\fancyhf{}\fancyfoot[C]{\normalsize\thepage}}
\fancypagestyle{firstpage}{\fancyhf{}\fancyfoot[C]{\normalsize\thepage}\fancyfoot[L]{\footnotesize Preprint.}}
\usepackage{enumitem}\setlist{itemsep=1pt,topsep=3pt,leftmargin=1.5em}

\usepackage{bookmark}
\IfFileExists{xurl.sty}{\usepackage{xurl}}{} 
\hypersetup{
  colorlinks=true,
  linkcolor={blue},
  filecolor={Maroon},
  citecolor={blue},
  urlcolor={blue},
  pdfcreator={LaTeX via pandoc}}

\author{}
\date{}

\begin{document}

\thispagestyle{firstpage}
\begin{center}
{\noindent\rule{\linewidth}{1.6pt}}\\[3pt]
{\LARGE\bfseries Measuring the Re-executability of Published\\[2pt]Molecular Docking Claims\par}
\vspace{5pt}
{\small\itshape MERS-Dock, an E0--E4 executability ladder, a 65-paper dual-human validation,\\and a bounded re-execution audit of the SARS-CoV-2 main-protease literature.\par}
\vspace{5pt}
{\noindent\rule{\linewidth}{0.5pt}}\\[13pt]
{\large\bfseries Vincent Giap\quad\quad Eric Wang\quad\quad Cris Nguyen\par}
\vspace{4pt}
{NewScience Lab\par}
\end{center}
\vspace{15pt}

\begin{center}{\large\bfseries Abstract}\end{center}

\begingroup\leftskip=2.4em \rightskip=2.4em \small

Published molecular docking scores are workflow outputs, not standalone
facts: they depend on the receptor, ligand, software, search box, random
seed, and preparation choices. We ask whether such claims can be
re-executed from their own published records. We introduce MERS-Dock, a
16-field Minimum Executable Reporting Set; a deterministic E0-E4 ladder
assigned over audited field states; and a re-execution protocol that
benchmarks observed deviations against AutoDock Vina's run-to-run noise
and a 2.0 kcal/mol operational tolerance. In 236 open-access,
text-extractable SARS-CoV-2 main-protease docking papers, 8.1\% met the
prespecified essential-field rule for direct re-execution (E3; 95\% CI
5.2-12.2), 44.1\% required explicit assumptions (E2), 47.9\% were
blocked by a missing foundational field (E1), and none reached E4
status; the paper-level distribution is byte-reproducible from the
released field-state matrix under the frozen rule. Mean field
completeness was 49.1\%; the search-box centre was reported by only
33.9\% of papers. We validated the automated field-state audit against
\textbf{two independent human reviewers on a 65-paper stratified sample}
(all directly-executable papers plus sampled blocked and
assumption-dependent ones): the reviewers agreed with each other on 92\%
of field states (pooled Cohen kappa 0.87), and the automated agent that
produced the corpus states agreed with humans on the execution-blocking
fields (PDB 89\%, software 92\%, box-centre prevalence 34 vs 32\%) while
systematically over-calling two non-blocking fields (numeric-result
reported 88\% by agent vs 37\% by humans; validation-redocking 42 vs
12\%). Because those over-called fields do not enter the executability
rule, the E-class was 68\% concordant between agent and humans and,
where it differed, human review lowered the executable count --- so the
low-executability headline is confirmed, not inflated, by human
validation. Reporting did not improve over five years (16-field
completeness vs publication year Spearman rho = -0.01, p = 0.92; no
execution-blocking field improved). As a bounded, mechanistic
re-execution, we re-docked one claim from each of 13 fully-reported
papers at its reported centre under three boxes --- its reported box, a
25 \AA{} default, and a ligand-volume-informed box. A reported box
reproduced the published score better than the default (median absolute
deviation 1.85 vs 2.51 kcal/mol; exact Wilcoxon signed-rank P = 0.027,
nine complete pairs), but the effect is fragile (it hinges on one paper;
P = 0.055 without it) and null for on-pocket focused-box papers (P =
0.44); a sensibly ligand-sized box did no better than the default (P =
0.20, an exploratory post-hoc arm), and the two papers no small box
could reproduce are exactly those whose reported centre lies 33--38
\AA{} off the pocket. The difference is therefore box-coverage geometry,
not box-size disclosure: reproducing a wide blind-docking result
requires the entire reported box, centre and size together. E-class acts
as an executability gate, not a reproducibility predictor. We release
Mpro-DockExec as a traceable measurement layer for digital-library and
evidence-synthesis systems deciding what is checkable in published
computational claims.

\par\endgroup
\vspace{6pt}

\textbf{Scope note.} This study audits the \textbf{open-access,
text-extractable} SARS-CoV-2 main-protease docking literature. It
measures reporting executability, validates the audit against
independent human reviewers on a stratified sample, and includes a
bounded re-execution proof of concept; it does not verify the literature
at scale.

\section{Significance}\label{significance}

A docking paper can report a score without reporting the workflow needed
to regenerate it. MERS-Dock turns that problem into a measurement: 16
execution-relevant fields, a deterministic E0-E4 label, and a deep-audit
ledger that records method and result spans at claim level. In 236
open-access Mpro docking papers, nearly half are execution-blocked by at
least one missing foundational field, and none meet the E4
robustness-ready rule --- a gap that has not narrowed in five years. An
independent two-reviewer human validation on a stratified 65-paper
sample confirms the audit is reliable (92\% inter-reviewer agreement)
and, importantly, shows the automated agent's largest errors fall on
fields that do not affect executability, so the low-executability
finding survives human scrutiny. The main fix is simple: disclose the
search box, configuration, and seed. A bounded re-execution then shows
the harder half of the problem is geometric: a published score is
regenerable only when the full search box --- centre and size together
--- is recoverable, and for wide blind-docking boxes placed at
off-pocket centres no standard reconstruction reaches the pocket, so an
unreported box is often irrecoverable. The result is a reusable
reporting and verification substrate: it tells a reader or automated
agent which claims can be re-executed cleanly before any scientific
correctness claim is made.

\section{Introduction}\label{introduction}

A molecular docking prediction depends jointly on the receptor structure
and chain, the ligand identity and protonation state, the software and
its version, the search-box geometry, the search effort, the random
seed, and the handling of waters and co-factors (Trott and Olson 2009;
Eberhardt et al. 2021; Guedes et al. 2013). Change any of these and the
score changes. A paper that reports only the score has not made a
testable claim; it has published a number with an unknowable provenance.
Mpro docking is an ideal place to measure this: the COVID-19 response
produced hundreds of Mpro virtual-screening studies in a few years
(Mandour et al. 2020; Sisakht et al. 2021; Ambrosio et al. 2023;
Peralta-Moreno et al. 2023).

The problem is recognised, and two prior audits have examined the
SARS-CoV-2 Mpro docking literature specifically --- but each measured
\emph{validation quality}, not \emph{re-executability}. Llanos and
colleagues found that a docking protocol was validated ``at some level''
in 57.7\% of 168 Mpro virtual-screening investigations, with only three
complete retrospective analyses (Llanos et al. 2021); a companion
critique of 61 Mpro repurposing manuscripts concluded that most do not
validate their methodology or their scores against activity (Macip et
al. 2021). These establish that the field under-validates. They do not
measure whether a paper's \emph{reported record} is sufficient to
regenerate its docking result, nor do they provide a machine-readable
schema or a computable label. More broadly, methodological audits have
flagged selective reporting in docking (Jain 2007), and large redocking
and scoring-function benchmarks show that performance is strongly
conditional on protocol choices that are often left implicit
(Flachsenberg et al. 2024; Zajaček et al. 2024; Warren et al. 2006;
Cross et al. 2009; Su et al. 2019). Recent guidance now specifies
practical reproducibility tips and minimum reporting items for docking
(Martis and Teletchea 2025; Kittelson et al. 2026), but it does not
provide a deterministic literature-audit instrument or estimate how
often published records satisfy those items. Across computational
biology, most published workflows cannot be regenerated from the
published record (Samuel and Mietchen 2023; Rule et al. 2019), part of a
reproducibility problem documented across science (Baker 2016; Sandve et
al. 2013; Peng 2011; Stodden et al. 2016). Adjacent computational fields
have answered with minimum-documentation standards, including FAIR data
principles (Wilkinson et al. 2016), datasheets (Gebru et al. 2021),
model cards (Mitchell et al. 2019), ARRIVE (Percie du Sert et al. 2020),
and MIAME (Brazma et al. 2001). The issue is now urgent for a new
reason: language models extract and propagate published claims at a
scale beyond human review (Gartlehner et al. 2024; Konet et al. 2024;
Yisha et al. 2026), and a claim that cannot be re-executed is one whose
errors cannot be caught before they spread.

What has been missing is an execution-oriented instrument that converts
published reporting into a computable label and an empirical
distribution. We introduce three coupled components. First, MERS-Dock, a
16-field Minimum Executable Reporting Set for molecular docking, tiered
by the role each field plays in execution and designed as an author
checklist, reviewer tool, and machine-readable schema for
scholarly-document audit. Second, an E0--E4 ladder assigned by a
deterministic rule over the audited field-state matrix, which we
validate against two independent human reviewers on a 65-paper
stratified sample. Third, a re-execution protocol that contextualises
score deviations with the docking engine's measured run-to-run noise
floor (median 0.04 kcal/mol) and a declared 2.0 kcal/mol operational
tolerance. We field-test the assembled method on 236 open-access,
text-extractable SARS-CoV-2 Mpro docking papers, localise which reported
fields most often block execution, and release the standard, classifier,
derived data, and scripts as Mpro-DockExec. Ranking claims by
trustworthiness and re-executing them at scale remain explicitly
downstream companion work.

\textbf{Contributions.} (1) MERS-Dock, an execution-oriented Minimum
Executable Reporting Set for molecular docking: 16 fields tiered by the
role each plays in execution (Tier-1 blocks execution when absent,
Tier-2 forces explicit assumptions, Tier-3 enables robustness checks).
It is released as a reusable author checklist, reviewer tool, and
machine-readable schema for digital-library audit. (2) An E0--E4
executability ladder with a deterministic decision rule assigned over
source-span-verified field states, reproducible from the released matrix
by construction and \textbf{validated against two independent human
reviewers on a 65-paper stratified sample} (inter-reviewer agreement
92\%, pooled Cohen kappa 0.87). The validation localises where the
automated agent is trustworthy (execution-blocking fields) and where it
is not (numeric-result and redocking, both non-blocking), and shows
human review does not raise --- and slightly lowers --- the executable
fraction. (3) A five-year temporal analysis showing execution-metadata
reporting has not improved (completeness vs year rho = -0.01, p = 0.92),
a result robust to a consistent agent bias because such a bias cancels
in a trend. (4) A bounded re-execution audit with an engine-noise
reference: deviations are read against Vina's measured self-noise
(median 0.04 kcal/mol) under a declared 2.0 kcal/mol operational
tolerance. Re-docking one claim from each of 13 fully-reported papers at
its reported centre under three boxes (reported, 25 \AA{} default, and a
post-hoc ligand-volume box) shows a reported box beats a 25 \AA{}
default (paper-level median \textbar{}\ensuremath{\Delta}\textbar{} 1.85
vs 2.51 kcal/mol; exact Wilcoxon p = 0.027, nine pairs), but the effect
is fragile (drop-one-paper p = 0.055; null for on-pocket focused boxes,
p = 0.44) and is not reproduced by a sensibly ligand-sized box (p = 0.20
vs the default) --- the difference is box-coverage geometry at
off-pocket blind-docking centres, not box-size disclosure, so the full
box (centre and size) is execution-critical. The executability label is
read as an executability gate, not a reproducibility predictor. (5) A
source-span-anchored, deterministically verified audit method, released
as Mpro-DockExec, demonstrated as a field-scale readout on 236
open-access Mpro papers. The search-box centre is the most common
execution-blocking omission (missing from two-thirds of papers; mean
completeness 49.1\%). Because the corpus is open-access and
text-extractable, the figures are an upper bound for that subset, not a
field-wide constant.

\section{Results}\label{results}

\subsection{A field-test corpus: open-access Mpro docking
papers}\label{a-field-test-corpus-open-access-mpro-docking-papers}

From 1,400 screened records (1,018 after deduplication), 241 had
retrievable open-access full text and 236 were confirmed original Mpro
docking studies (Table 1; PRISMA 2020 flow, Fig. S1). These papers carry
roughly 12,466 ligand-level claim rows, but a single high-throughput
screen contributes \textasciitilde9,600 of them, so we treat that total
as a rough upper bound rather than a count of independent docking claims
and report \textbf{paper-level} statistics as primary throughout. The
corpus is therefore restricted to open-access, text-extractable
literature: every audited paper was processed from genuine full text
(PMC JATS, publisher open-access, preprints), never from a paywalled or
PDF-only route. This scope is a deliberate strength for reliability and
an explicit limitation for generality, addressed in the Discussion.

\subsection{The method: MERS-Dock, a deterministic executability ladder,
and a computable
label}\label{the-method-mers-dock-a-deterministic-executability-ladder-and-a-computable-label}

We define MERS-Dock, the Minimum Executable Reporting Set for docking:
16 fields grouped by execution role. \textbf{Tier-1} fields block
execution when absent: the receptor/PDB, the search-box centre, and the
docking software. \textbf{Tier-2} fields do not block execution but
force assumptions when absent, and govern the step from
assumption-dependent (E2) to directly executable (E3): box size,
software version, search effort, receptor and ligand preparation,
protonation state, water/ion handling, protein chain, and ligand
identity. \textbf{Tier-3} fields enable robustness checks: random seed,
validation redocking, reusable configuration. The deterministic E-class
rule below uses exactly this structure: a missing Tier-1 field forces
E1, full Tier-1-plus-essential reporting reaches E3, and the disclosed
seed and configuration of Tier-3 mark E4.

Each paper is placed on a five-level ladder by a deterministic rule over
the verified field states (Fig. 1). Define the blocking set, essential
set, and a shorthand for complete reporting as

\[
\begin{aligned}
\mathrm{BLOCK}&=\{\mathrm{pdb},\mathrm{grid\_centre},\mathrm{software}\},\\
\mathrm{ESS}&=\{\mathrm{pdb},\mathrm{grid\_centre},\mathrm{grid\_size},\mathrm{software}\}\\
&\quad\cup\{\mathrm{version},\mathrm{receptor\_prep},\mathrm{ligand\_prep}\},\\
R(S)&\equiv\forall f\in S:\ \mathrm{state}(f)=\mathrm{reported}.
\end{aligned}
\]

\[
E(\mathrm{paper})=\begin{cases}
E_1 & \exists f\in \mathrm{BLOCK}:\ \mathrm{state}(f)=\mathrm{missing}\\
E_4 & R(T_1\cup T_2)\wedge \mathrm{seed}\wedge \mathrm{config}\\
E_3 & R(\mathrm{ESS})\wedge \mathrm{ligand\_id}\ne\mathrm{missing}\\
E_2 & \mathrm{otherwise}.
\end{cases}
\]

E-class describes the reporting record, not the science: an E2 result
may be entirely correct, it simply does not uniquely determine its
workflow. Scoring is strict --- only \texttt{reported} counts;
\texttt{partial} is treated as not-reported --- and the label is
reproducible from the released field states by construction (the
236-paper distribution below re-derives byte-for-byte from the matrix
under this rule).

\begin{figure}
\centering
\pandocbounded{\includegraphics[keepaspectratio,alt={Figure 1. MERS-Dock maps three tiers of reporting fields to a deterministic E1-E4 executability ladder. Tier-1 omissions block execution, Tier-2 omissions force explicit assumptions, and Tier-3 fields enable full reproducibility and robustness checks.}]{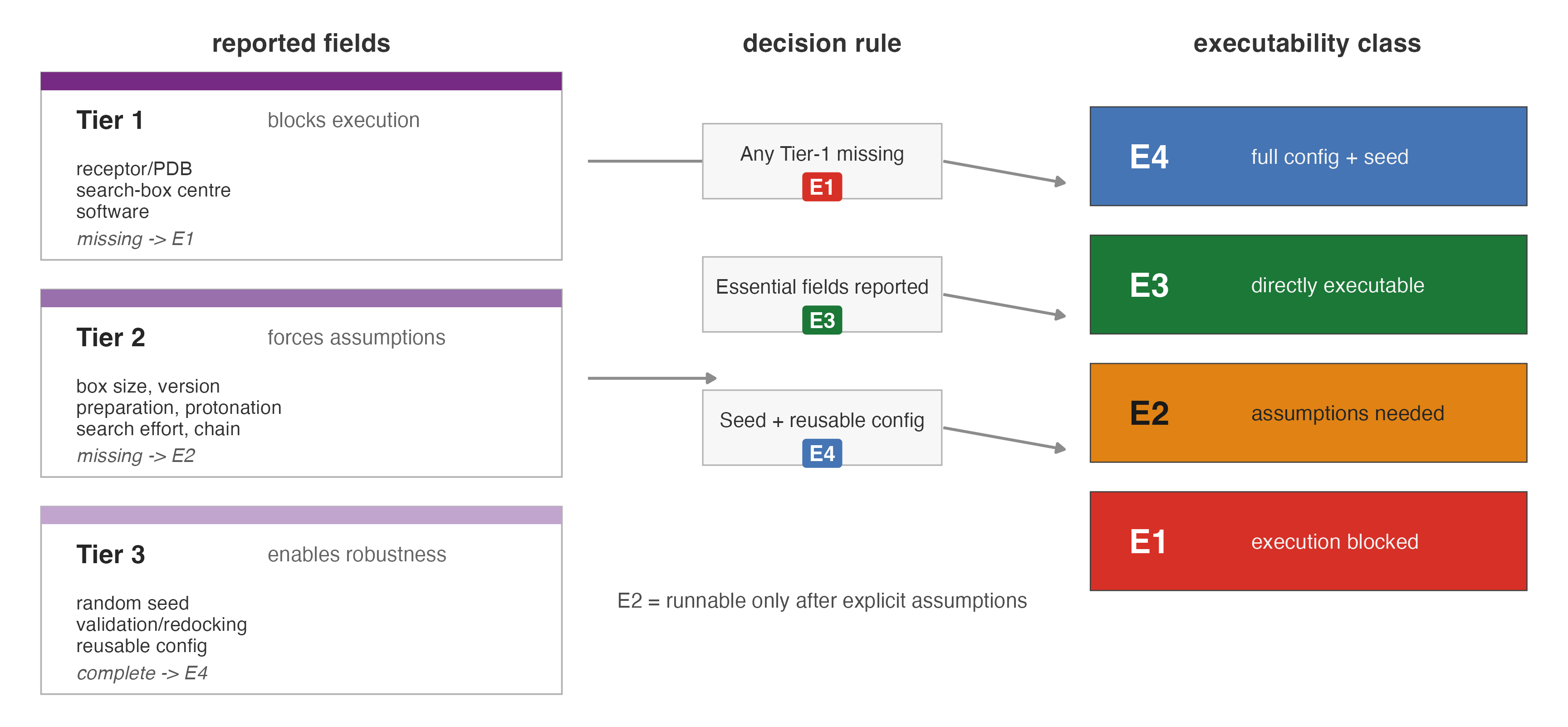}}
\caption{Figure 1. MERS-Dock maps three tiers of reporting fields to a
deterministic E1-E4 executability ladder. Tier-1 omissions block
execution, Tier-2 omissions force explicit assumptions, and Tier-3
fields enable full reproducibility and robustness checks.}
\end{figure}

\subsection{Applying the standard: which fields are reported, and which
block
execution}\label{applying-the-standard-which-fields-are-reported-and-which-block-execution}

Applying MERS-Dock across the corpus localizes the reporting gap to
specific execution-critical fields rather than to reporting in general.
Only two fields were near-universal: docking software (97.9\%) and
receptor PDB (94.9\%). Completeness then fell sharply, and it fell
precisely where re-execution depends most (Table 2, Fig. 2). The
search-box centre, the field that most often blocks execution, was
reported in only 33.9\% of papers; both box parameters appeared together
in only 27.5\% (65/236). The random seed, the cheapest field to deposit
and the one that enables exact reproduction, was disclosed by a single
paper. Rather than average all sixteen fields into one number (which
would weight a PDB identifier and a random seed equally), we read
completeness by execution tier, and it collapses monotonically down the
tiers that matter for re-execution: the execution-blocking Tier-1 fields
are reported 75.6\% of the time, the assumption-forcing Tier-2 fields
47.5\%, and the robustness Tier-3 fields --- seed, configuration,
validation --- only 12.3\% (the overall 16-field mean is 49.1\%). The
fields that make a result re-executable and checkable are precisely the
least reported.

\begin{figure}
\centering
\pandocbounded{\includegraphics[keepaspectratio,alt={Figure 2. Reporting completeness across 16 MERS-Dock fields (N = 236; Wilson 95\% CI), ranked by reporting frequency and coloured by execution tier. The search-box centre is reported by only 33.9\% of papers.}]{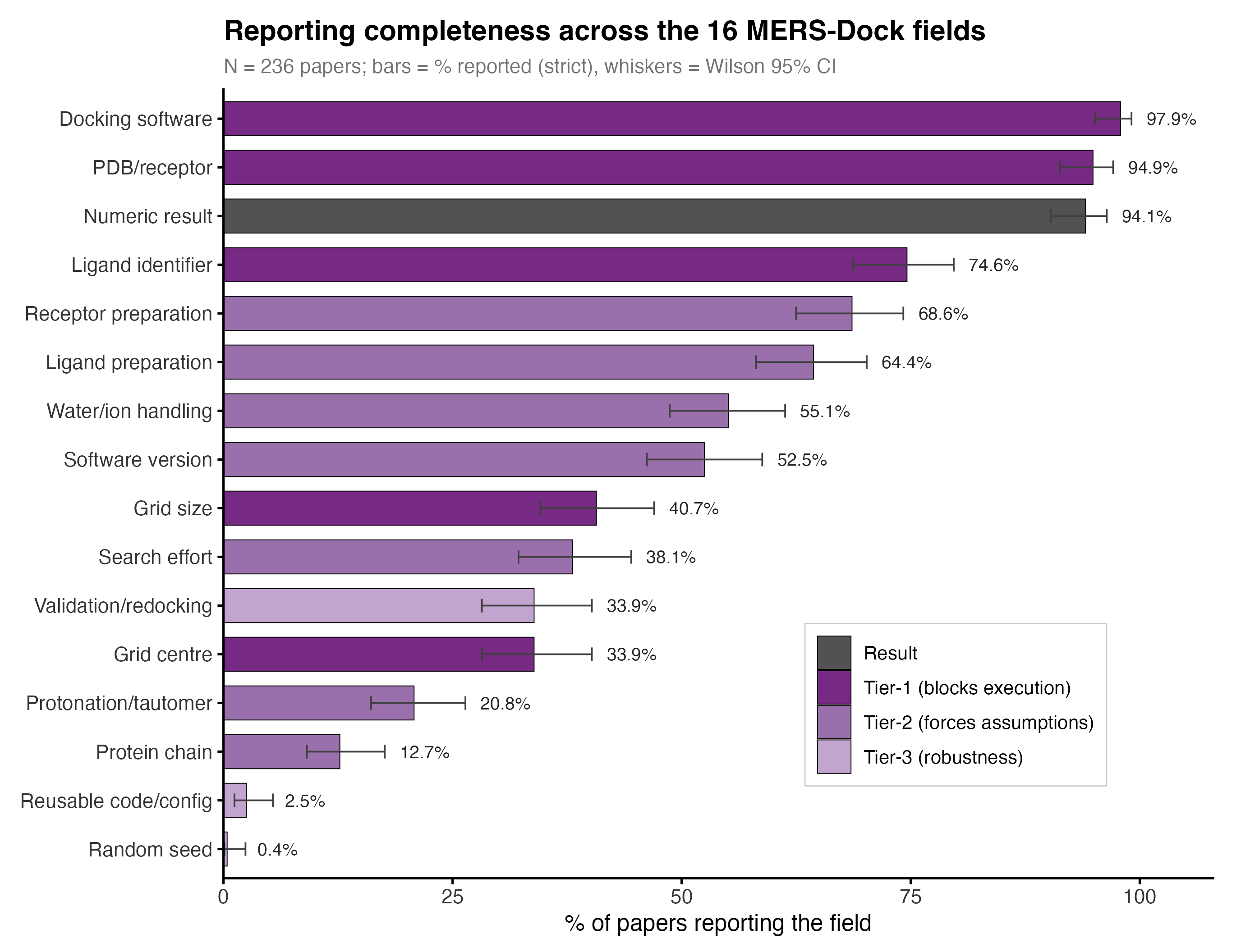}}
\caption{Figure 2. Reporting completeness across 16 MERS-Dock fields (N
= 236; Wilson 95\% CI), ranked by reporting frequency and coloured by
execution tier. The search-box centre is reported by only 33.9\% of
papers.}
\end{figure}

\begin{figure}
\centering
\pandocbounded{\includegraphics[keepaspectratio,alt={Figure 3. Per-paper missing-parameter matrix (236 papers x 16 fields; blue reported, amber partial, red missing), ordered by E-class and completeness. The execution-blocked block is dominated by missing search-box centres.}]{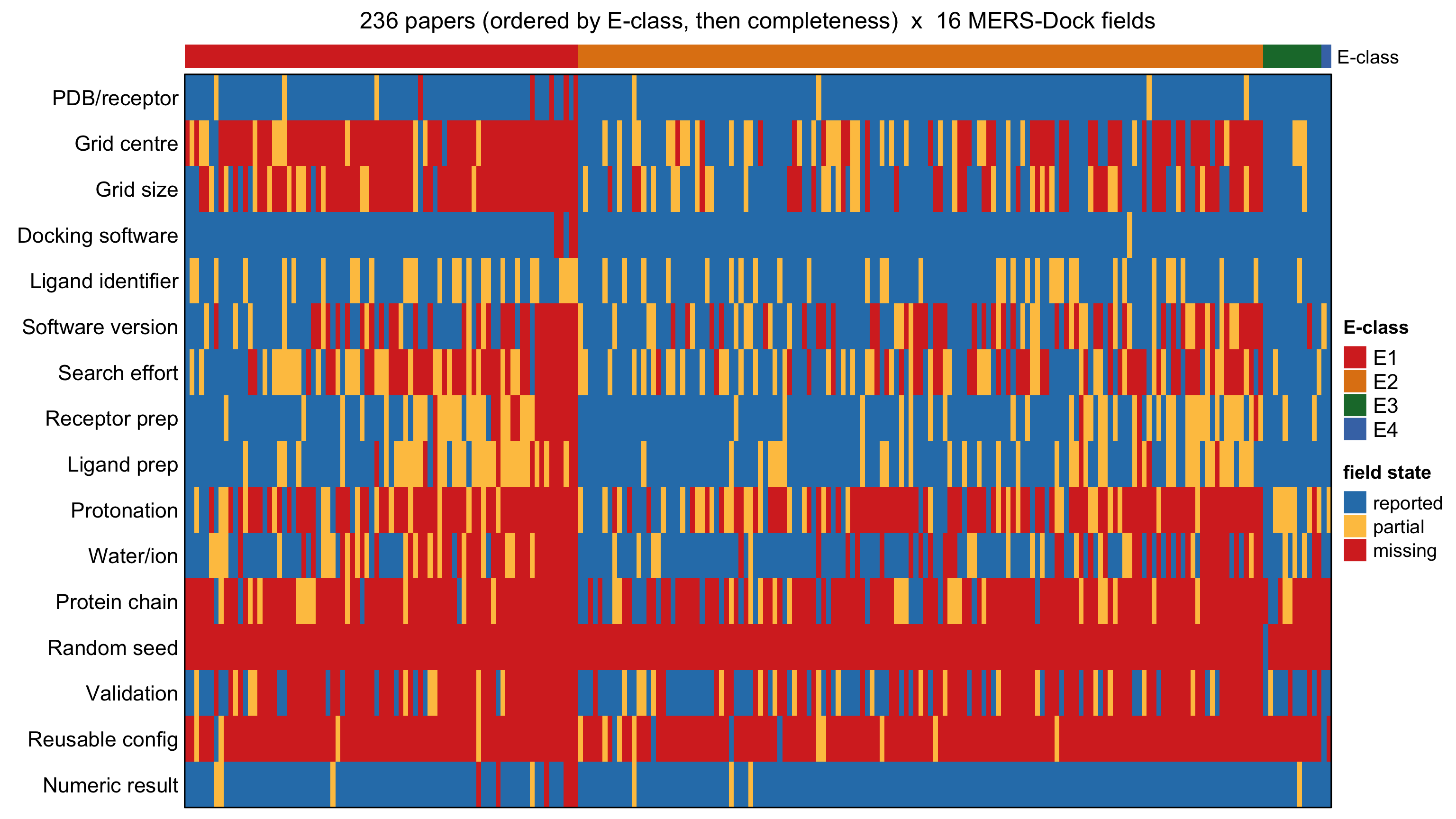}}
\caption{Figure 3. Per-paper missing-parameter matrix (236 papers x 16
fields; blue reported, amber partial, red missing), ordered by E-class
and completeness. The execution-blocked block is dominated by missing
search-box centres.}
\end{figure}

\subsection{The executability label across the corpus (E1--E4), and what
it localizes to
fix}\label{the-executability-label-across-the-corpus-e1e4-and-what-it-localizes-to-fix}

Assigned by its deterministic rule at the paper level (N = 236, the
manuscript's main endpoint after a logged post hoc unit-of-analysis
deviation; Table 3, Fig. 4), the executability ladder resolves a clear
gradient: 47.9\% of papers are execution-blocked (E1; 95\% CI
41.6--54.2), 44.1\% are executable only after explicit assumptions (E2;
37.9--50.4), 8.1\% are directly re-executable (E3; 5.2--12.2), and none
reach full reproducibility (E4; 0/236, 0.0--1.6). No paper was E0, the
floor class for a record with no extractable docking claim. The label
thus localizes what to fix: directly executable papers are about one in
twelve, and nearly half cannot be run at all because a single Tier-1
field is missing. This distribution is a frozen-rule recomputation over
the released field states and reproduces exactly.

No paper achieved E4: the one paper that reports a seed omits its
configuration. Full reproducibility is achievable with only a few extra
disclosed fields, but in this corpus no one achieves it. At the
claim-weighted level the distribution is E1 10.6\%, E2 86.9\%, E3 2.4\%,
E4 0\%, but this is dominated by one high-throughput screen
(\textasciitilde77\% of all claim rows, \textasciitilde88\% of E2
claims): removing that single paper moves the claim-weighted
distribution to E1 45.6\% / E2 43.9\% / E3 10.5\%, close to the
paper-level figures. Completeness does not increase with output
(Spearman \ensuremath{\rho} \ensuremath{\approx} 0.04, n.s.), so the
claim-weighted concentration reflects one bulk paper, not a reporting
tendency; the paper-level distribution is primary.

The directly-executable fraction depends on how strict the E3 bar is.
Across an exhaustive sweep of rule variants scored against the reference
labels, the execution-blocked fraction is highly stable (E1
47.5--47.9\%), whereas the directly-executable fraction ranges
1.3--8.1\% across the principal reference-aligned rules --- all of which
score fields under strict reported-only coding. Relaxing that coding so
partially-reported essential fields count as adequate raises E3 to about
30\%, so the 8.1\% figure is conditional on strict scoring; only the
blocked-majority finding (E1 \textasciitilde48\%) is robust to the
coding choice.

\begin{figure}
\centering
\pandocbounded{\includegraphics[keepaspectratio,alt={Figure 4. E1-E4 executability distribution with Wilson 95\% intervals at paper level (N = 236) and claim-weighted level (\textasciitilde12,466 claims). Labels come from the deterministic rule classifier.}]{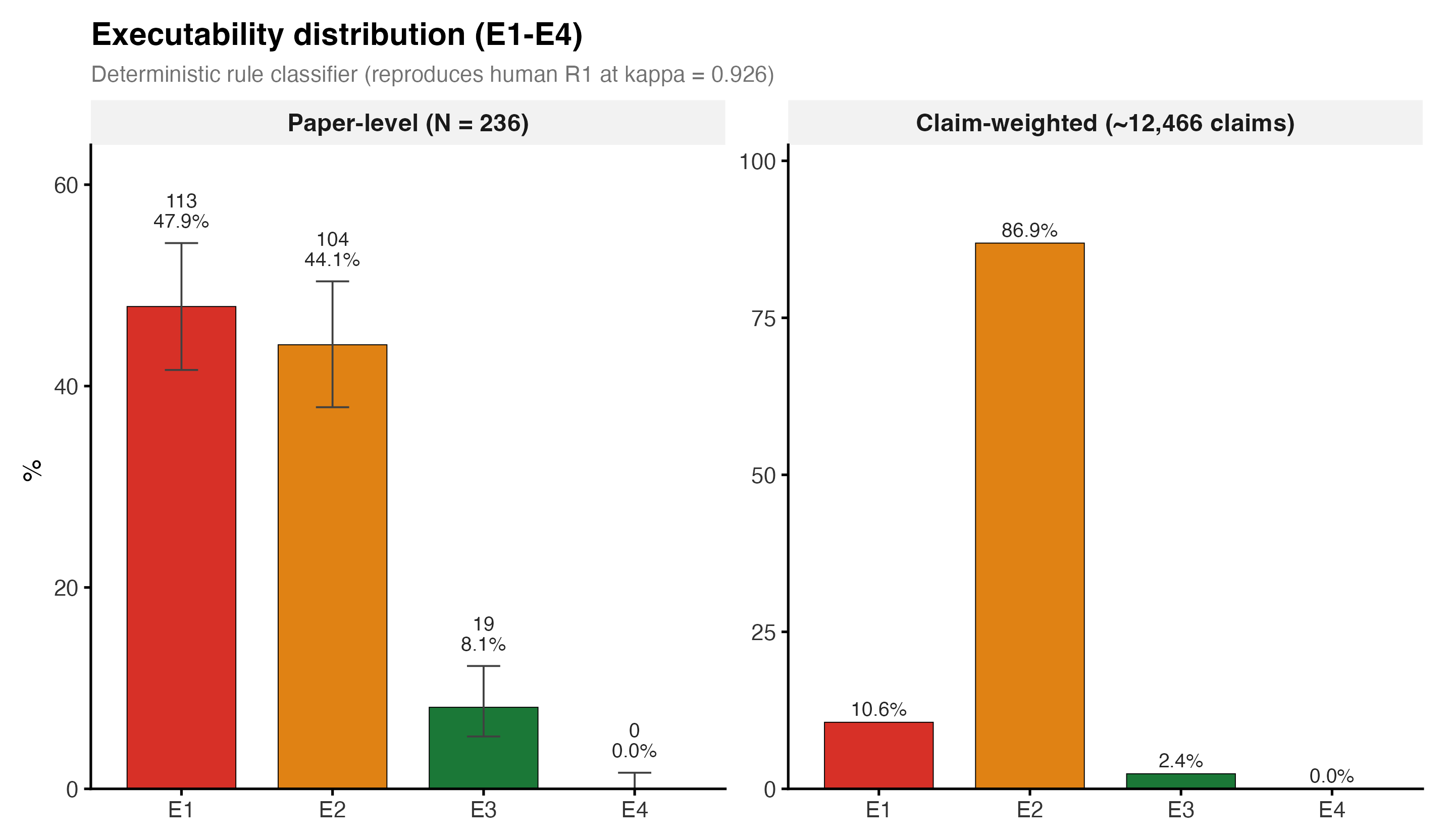}}
\caption{Figure 4. E1-E4 executability distribution with Wilson 95\%
intervals at paper level (N = 236) and claim-weighted level
(\textasciitilde12,466 claims). Labels come from the deterministic rule
classifier.}
\end{figure}

\subsection{Validating the audit: two independent human reviewers on 65
papers}\label{validating-the-audit-two-independent-human-reviewers-on-65-papers}

The audit's field states are produced by an automated extraction agent,
so the load-bearing question is whether those states are correct. We
answered it with a prospective, two-reviewer human validation. From the
236-paper corpus we drew a \textbf{stratified 65-paper sample} --- every
directly-executable (E3) and E4-candidate paper plus a random sample of
blocked (E1) and assumption-dependent (E2) papers (sampling manifest
released) --- and had \textbf{two independent human reviewers} re-assess
all 16 MERS-Dock field states for each paper from the paper's own text,
blind to each other, with the extraction agent's values shown only as a
checklist to accept or overrule. Disagreements were adjudicated by a
third human. All E-classes in this section are recomputed from the
human-adjudicated states with the identical frozen rule used for the
corpus (Fig. 5).

The two human reviewers agreed on \textbf{92\% of the 1,040 field
states} (pooled Cohen kappa 0.87 --- ``almost perfect'' on the
Landis--Koch scale). Per-field kappa was \ensuremath{\geq} 0.80 for
eleven of sixteen fields, including every execution-blocking one
(docking software 1.00, grid-centre 0.86; the receptor-PDB field reads
kappa 0.56 only because it is almost always reported --- a prevalence
artefact for which the prevalence-robust Gwet AC1 is 0.97 at 92\%
agreement). This is the independent human inter-rater reliability the
corpus previously lacked, and it certifies that the field states --- not
merely the codebook --- are reproducibly judged (Table 4b).

The validation also prices the automated agent. On the fields that
determine executability, the agent tracked the humans closely: it
matched the human on the receptor-PDB 89\% of the time and on the
docking software 92\%, and its box-centre \emph{prevalence} (34\%)
essentially equalled the human value (32\%). But it
\textbf{systematically over-called two non-execution fields}: it marked
a usable numeric result as \texttt{reported} in 88\% of papers where
humans found only 37\% (a 52\% per-paper disagreement), and marked
redocking validation \texttt{reported} in 42\% versus the human 12\%
(Fig. 5B). Crucially, neither field enters the E-class rule ---
\texttt{numeric\_result} is not in the ladder and
\texttt{validation\_redocking} is a Tier-3 robustness field that affects
only the (empty) E4 class. Because the agent's largest errors miss the
executability-determining fields, the \textbf{E-class agreed between
agent and humans on 68\% of the 65 papers}, and where it differed the
net effect of human review was to \emph{lower} the executable count
(agent E3 = 10 \ensuremath{\rightarrow} human E3 = 8), with the
remaining changes being E1\ensuremath{\leftrightarrow}E2 boundary churn
that nearly cancels (Fig. 5C). Human validation therefore confirms the
headline rather than inflating it: the true executable fraction is, if
anything, slightly below the agent's estimate, and the over-reported
fields are ones a downstream user must human-verify regardless (see
Limitations).

\begin{figure}
\centering
\pandocbounded{\includegraphics[keepaspectratio,alt={Figure 5. Sixty-five-paper dual-human validation of the automated field-state audit (frozen MERS-Dock v1.1 rule). (A) Two independent human reviewers agree on 92\% of field states (per-field Cohen kappa; pooled 0.87); PDB kappa=0.56 is a prevalence paradox (Gwet AC1=0.97). (B) The automated agent (open dot) vs human consensus (filled) reported-rate per field, ranked by disagreement: the agent over-calls numeric\_result (+51 pp) and validation-redocking (+30 pp) --- both non-blocking --- while matching humans on execution-blocking fields. (C) E-class from agent vs human states (68\% concordant); human review lowers the executable count (E3 108).}]{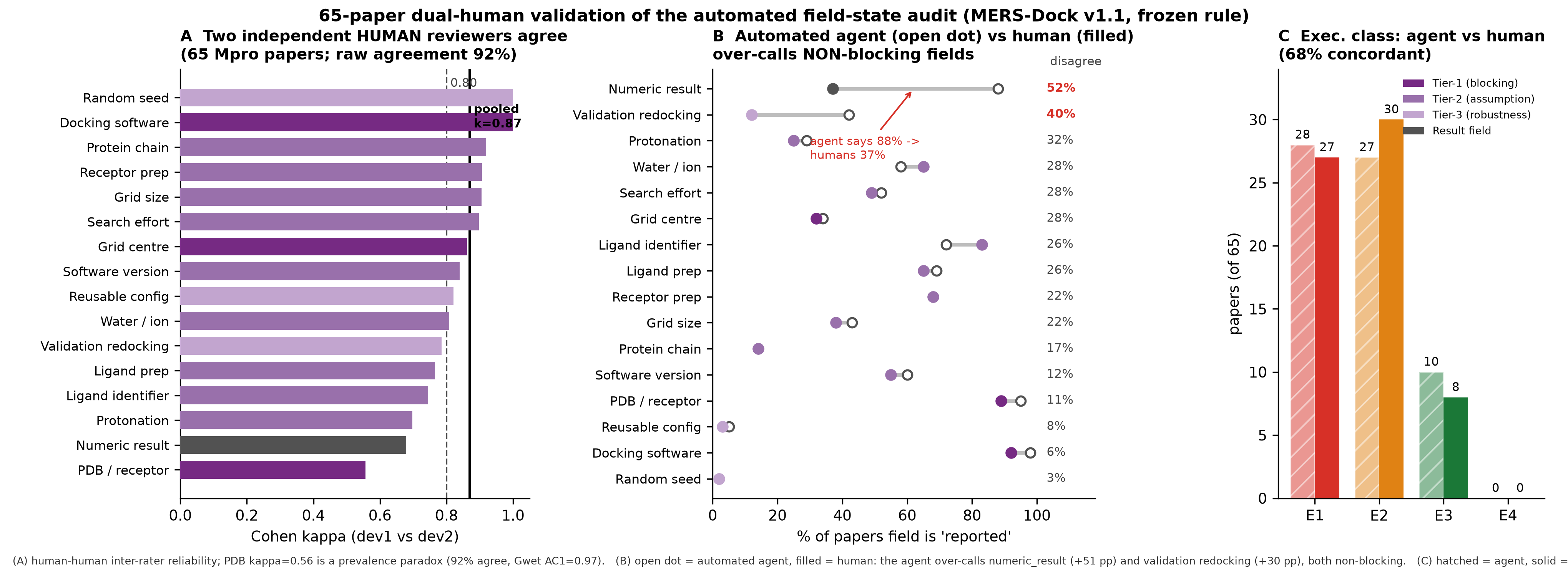}}
\caption{Figure 5. Sixty-five-paper dual-human validation of the
automated field-state audit (frozen MERS-Dock v1.1 rule). (A) Two
independent human reviewers agree on 92\% of field states (per-field
Cohen kappa; pooled 0.87); PDB kappa=0.56 is a prevalence paradox (Gwet
AC1=0.97). (B) The automated agent (open dot) vs human consensus
(filled) reported-rate per field, ranked by disagreement: the agent
over-calls numeric\_result (+51 pp) and validation-redocking (+30 pp)
--- both non-blocking --- while matching humans on execution-blocking
fields. (C) E-class from agent vs human states (68\% concordant); human
review lowers the executable count (E3 10\ensuremath{\rightarrow}8).}
\end{figure}

\subsection{Reporting has not improved over five
years}\label{reporting-has-not-improved-over-five-years}

Because a single consistent agent bias cancels in a temporal
\emph{trend} even if it shifts the \emph{level}, the change in reporting
over time is the most bias-robust readout in the corpus. It shows
\textbf{no improvement}. Across the 236 papers spanning 2021--2026,
16-field completeness versus publication year has Spearman rho = -0.01
(p = 0.92): flat. None of the execution-blocking fields improved ---
grid-centre rho = -0.09 (p = 0.15), grid-size -0.08 (p = 0.21),
random-seed +0.08 (p = 0.22, and it stays near 0\%), reusable-config
-0.11 (p = 0.09) --- and the flat trend is not a composition artefact,
because recent papers are \emph{smaller} screens, not bulk ones
(claims-per-paper vs year rho = -0.35, p \textless{} 0.001), and bulk
papers report the grid-centre slightly more often, not less. Only two
downstream \emph{practice} fields rose (redocking validation rho =
+0.16, p = 0.013; protonation +0.13, p = 0.05). Reporting guidelines and
reproducibility checklists published across this window (Martis and
Teletchea 2025; Kittelson et al. 2026) have not moved the fields that
block re-execution. We state this as \emph{flat / no improvement}, not a
decline: a period table can look like a grid-centre drop
(41\%\ensuremath{\rightarrow}27\%), but the monotone trend test is not
significant.

\subsection{Re-executing an unreported search box: a coverage-geometry
problem, not a box-size
effect}\label{re-executing-an-unreported-search-box-a-coverage-geometry-problem-not-a-box-size-effect}

Whether a published score can be regenerated is only cleanly testable on
the reported subset, because re-running an under-reported claim requires
supplying the very field whose absence defines it. We therefore designed
the primary re-execution test as a within-paper paired comparison. A
two-arm protocol --- the paper's own \textbf{reported box} versus a
standardized \textbf{25 \AA{} default box}, both at the identical
reported centre --- was frozen before the runs were inspected (protocol
frozen 2026-07-11); a third \textbf{ligand-volume-informed box}
(isotropic edge = ligand heavy-atom diameter + 10 \AA{}, clamped to
22.5--40 \AA{}) was added \textbf{post hoc, after the two-arm result was
known}, as an exploratory control for whether a \emph{sensible size
guess} recovers the score, and is labelled exploratory throughout. The
unit of analysis is one paper. From every open-access paper reporting a
complete AutoDock Vina Mpro run --- receptor PDB, box centre, box size,
an identifiable ligand, and a numeric score, each anchored to a verified
source span --- we selected one claim before docking and re-docked it
under the three conditions, holding receptor, ligand, engine (AutoDock
Vina 1.2.7), search effort, and three deterministic seeds (11, 29, 53)
fixed across arms. Nineteen papers passed the span-verified eligibility
gate. One was excluded when its non-specific ``compound 17'' label
mis-matched an unrelated iodinated contrast agent, and five more had no
ligand resolvable to a public structure, leaving thirteen with a docked
structure (PubChem, CAS or ZINC identity, each confirmed by InChIKey;
ivermectin, returned as a homolog mixture, was docked as its major
component). Of these, one produced no bound pose under either box and
drops out, so twelve papers contributed a pose in at least one arm and
nine produced a pose under both boxes --- the nine complete pairs that
carry the primary analysis (full claim- and paper-level attrition in
Fig. S3). Each paper is thus its own control, which removes the
confounding that limits a between-paper comparison.

Three findings together show that the difference between reconstructions
is a matter of pocket \textbf{coverage}, not of box size and not of the
mere availability of the box. First, the reported box did reproduce the
published score better than the 25 \AA{} default (paper-level median
absolute deviation 1.85 vs 2.51 kcal/mol; paired median benefit 0.52;
exact Wilcoxon signed-rank \emph{P} = 0.027, nine complete pairs;
20,000-resample bootstrap 95\% CI of the benefit -0.005 to 0.96,
i.e.~touching zero), but that effect is fragile: it hangs on a single
paper --- dropping the one paper with an off-pocket centre (index 481)
moves \emph{P} to 0.055 --- and it vanishes in the clean regime, where
for the five papers whose box is focused (\ensuremath{\leq} 40 \AA{})
and whose centre lies on the pocket the paired benefit is 0.04 kcal/mol
(\emph{P} = 0.44). Second, the benefit scales with the reported box size
(Spearman \ensuremath{\rho} = 0.80, \emph{P} = 0.009): it is confined to
four wide blind-docking boxes (60--72 \AA{}) and is absent for focused
boxes. Third, and decisively, the exploratory
\textbf{ligand-volume-informed box --- the box a re-executor would
sensibly choose --- did no better than the naive 25 \AA{} default}
(median difference 0.02 kcal/mol, \emph{P} = 0.20) and was statistically
indistinguishable from the reported box (\emph{P} = 0.30). Had the
effect been about box \emph{size}, a correctly sized box should have
recovered the reported score; it did not. The engine is not the source
of the spread (within-paper seed range median 0.058, max 0.82 kcal/mol;
AutoDock Vina is near-deterministic).

The mechanism is geometric. The papers whose scores no small box could
regenerate are those whose reported centre lies far from the actual
binding pocket: measured against the co-crystallised-ligand centroid,
paper 59's reported centre sits 33 \AA{} from the 6LU7 pocket and paper
324's 38 \AA{}, so a 25 \AA{} cube --- or a 30 \AA{} ligand box ---
placed at those centres (half-width 12--15 \AA{}) physically cannot
reach the site and returns no bound pose under either standardized box.
(Paper 253, whose centre lies 17 \AA{} off-pocket, is a boundary case:
it failed the 25 \AA{} default but \emph{did} dock under the
ligand-volume box in all three seeds, so it is not in the no-pose set.)
The McNemar test on the ``reported-box-only pose'' pattern is not
significant (\emph{P} = 0.25). In these blind-docking studies the
reported centre alone does not locate the pocket; only the paper's own
wide box happened to span it. Reproducing such a claim therefore
requires the \emph{entire} reported box geometry --- centre and size
together --- not the centre plus any sensible size guess (Fig. 6). This
is a bounded, mechanistic result on a small, difficult subset, not
evidence that disclosing the box size per se improves reproduction.

A larger claim-level case series points the same way but is not an
independent test. Re-docking every curated ligand claim yielded 48
quality-controlled claims across three targets (6LU7, 6Y2E, 6yb7); those
run with a reported or capped box size reproduced to a median of 0.33
kcal/mol (11/12 within tolerance) versus 0.91 for the 36 forced onto a
25 \AA{} default (28/36; Mann--Whitney \emph{P} = 0.024; Table 5, Fig.
S2; per-claim CID/InChIKey scores in Table S1). These 48 claims are
clustered in only eight papers and --- unlike the paired design --- no
single paper carried both box conditions, so box availability is
confounded with paper identity and the comparison is pseudoreplicated.
Collapsed to those eight source papers, the contrast stays directional
(paper-level median 0.34 vs 1.15 kcal/mol) but is \emph{not} significant
(exact Mann--Whitney P = 0.20, four vs four papers): the claim-level P =
0.024 is inflated by pseudoreplication. We therefore report the series
only as exploratory, direction-consistent support; the paired
within-paper design, not this case series, carries the inference. The
executability class co-varies with the same signal (E3 median 0.24, n =
6, vs E2 0.83, n = 42) but is confounded with box-size source and rests
on an E3 stratum of six claims drawn largely from one paper (P0006;
\emph{P} = 0.07), so it is not an independent predictor.
Execution-blocked (E1) papers, lacking a box centre, cannot be
reconstructed without inventing one, which we did not do; the
re-execution set spans only E2 and E3.

We therefore read the executability label as an executability gate, not
a reproducibility score: it identifies which claims can be re-executed
at all --- the precondition an automated agent needs before deciding
whether re-execution is even an applicable verification tool. And the
re-execution itself yields a narrower, mechanistic lesson rather than a
headline effect: a published docking score is regenerable only when the
\emph{complete} search geometry is recoverable, and for the wide
blind-docking boxes at off-pocket centres in our eligible set, neither
the disclosed centre nor a sensibly sized box guess suffices --- the
full box, centre and size together, is execution-critical. This is what
justifies MERS-Dock treating both box fields as required; it is not a
claim that box-size disclosure alone improves regeneration. A clean
re-execution is in any case only ever possible on the reported subset,
because re-running an under-reported claim requires supplying the very
field whose absence defines it --- a boundary intrinsic to the approach.

\begin{figure}
\centering
\pandocbounded{\includegraphics[keepaspectratio,alt={Figure 6. Re-execution difference is a coverage-geometry effect, not a box-size effect (nine complete pairs of thirteen docked papers; AutoDock Vina 1.2.7, three seeds; full attrition Fig. S3). Each claim is re-docked at its reported centre under three boxes --- its reported box, a 25  default, and a post-hoc ligand-volume box. (A) the paired benefit (default  reported) scales with reported box size (Spearman  = 0.80, P = 0.009); labels give each paper's reported-centre-to-pocket offset. (B) per-paper median deviation for all three boxes, sorted by pocket offset: a ligand-volume box (orange) tracks the 25  default (red), not the reported box (green) --- volume vs default P = 0.20, reported vs volume P = 0.30 (both n.s.); off-pocket papers 59/324 (33--38  off-pocket) yield no pose under any small box. (C) the ``reported box helps'' effect is fragile: only the full nine-pair test clears 0.05 (P = 0.027), and it hinges on one off-pocket paper (481  P = 0.055); it is null for on-pocket focused-box papers (P = 0.44). The exploratory 48-claim case series is Fig. S2.}]{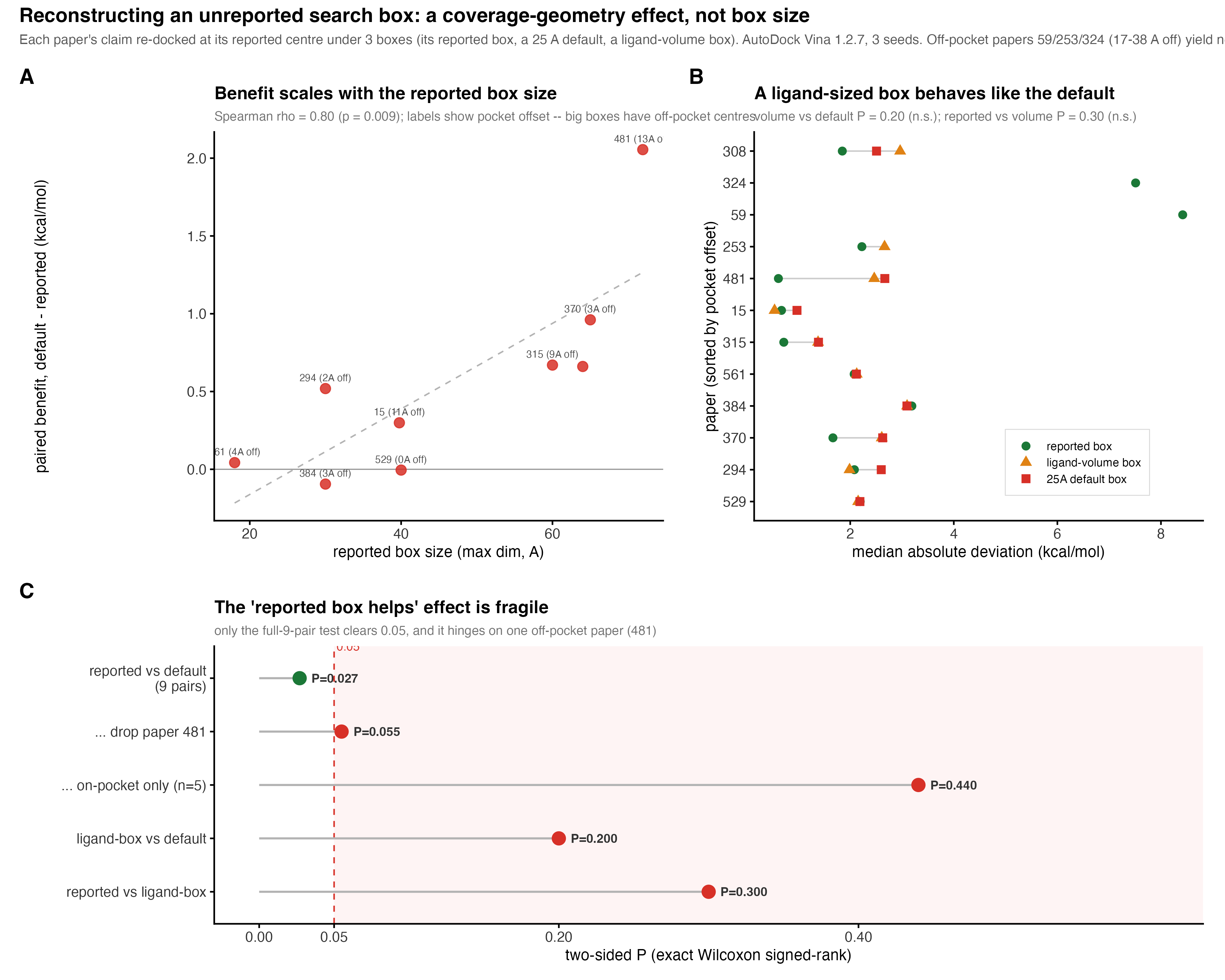}}
\caption{Figure 6. Re-execution difference is a coverage-geometry
effect, not a box-size effect (nine complete pairs of thirteen docked
papers; AutoDock Vina 1.2.7, three seeds; full attrition Fig. S3). Each
claim is re-docked at its reported centre under three boxes --- its
reported box, a 25 \AA{} default, and a post-hoc ligand-volume box. (A)
the paired benefit (default \ensuremath{-} reported) scales with
reported box size (Spearman \ensuremath{\rho} = 0.80, P = 0.009); labels
give each paper's reported-centre-to-pocket offset. (B) per-paper median
deviation for all three boxes, sorted by pocket offset: a ligand-volume
box (orange) tracks the 25 \AA{} default (red), not the reported box
(green) --- volume vs default P = 0.20, reported vs volume P = 0.30
(both n.s.); off-pocket papers 59/324 (33--38 \AA{} off-pocket) yield no
pose under any small box. (C) the ``reported box helps'' effect is
fragile: only the full nine-pair test clears 0.05 (P = 0.027), and it
hinges on one off-pocket paper (481 \ensuremath{\rightarrow} P = 0.055);
it is null for on-pocket focused-box papers (P = 0.44). The exploratory
48-claim case series is Fig. S2.}
\end{figure}

\subsection{The Mpro-DockExec release: standard, classifier, and
re-execution scripts as a measurement
layer}\label{the-mpro-dockexec-release-standard-classifier-and-re-execution-scripts-as-a-measurement-layer}

We release Mpro-DockExec: the 236-paper corpus with paper-level E-class
and 16-field states; a span-free 283-claim table carrying method/result
span-verification flags; the 65-paper human-validation package
(per-paper dual-reviewer field states, adjudicated states, per-field
agreement/kappa, sampling manifest, and reviewer attestation); the
reporting standard; the reproducible analysis and re-execution scripts,
including the frozen two-arm paired protocol, per-paper paired inputs
(receptor/ligand SHA-256, resolved SMILES/InChIKey/CID) and the
three-box docking outputs (\texttt{paired\_manifest.csv},
\texttt{input\_provenance.csv}, \texttt{paired\_runs.csv},
\texttt{volume\_runs.csv}) and locked statistics with pocket-offset
geometry (\texttt{reexec\_v2\_paper.csv},
\texttt{reexec\_v2\_stats.json}) alongside the exploratory case series
(\texttt{reproduction\_outcomes.csv}, \texttt{self\_consistency.csv});
and an RO-Crate provenance package. The dataset is a measurement layer
for scholarly records: it records what each paper reports and therefore
what is checkable. It does not re-run the literature at scale.
Copyrighted full texts and verbatim evidence spans are not
redistributed; derived metadata and span-verification flags are
released.

\section{Discussion}\label{discussion}

MERS-Dock and the executability ladder give the field the instrument it
lacked: a way to measure, not just assert, whether a published docking
score can be re-run, and to localize exactly what blocks re-execution.
Prior audits of this same Mpro literature measured whether protocols
were \emph{validated} and whether scores \emph{predict activity} (Llanos
et al. 2021; Macip et al. 2021); MERS-Dock instead measures whether the
\emph{reported record suffices to regenerate the result}, and turns that
into a computable label with an independent human-validated reliability.
Applied here, the instrument shows that the gap is a
workflow-description problem, not a data-sharing problem --- most papers
deposit figures and numbers but omit the parameters that determine the
score --- and it prices the remedy. The fix is almost free: a single
configuration file and one disclosed seed would move most E2 papers to
E3 or E4, yet no paper deposits both (E4 = 0/236) and about 97\% report
neither a reusable configuration (2.5\%) nor a seed (0.4\%). The
standard, the infrastructure and the near-zero cost all exist; the
measurement layer we release supplies the one missing piece --- a field
norm, and a tool to enforce it, that require them. And the gap is not
closing on its own: over five years, execution-metadata reporting is
flat (rho = -0.01), so a passive expectation that the field will
self-correct is not supported by the data.

\textbf{The audit is human-validated, and its errors are localized.} Two
independent human reviewers agreed on 92\% of field states (pooled kappa
0.87), giving the corpus the independent inter-rater reliability that a
codebook-only reproducibility argument cannot. The same validation shows
the automated agent is trustworthy exactly where it matters --- the
execution-blocking fields that drive E-class --- and untrustworthy on
two non-blocking fields it over-reports (numeric result, redocking
validation). This has a practical consequence: the executability
headline is safe to read at scale, but any \emph{specific} claim about
whether a paper reports a usable numeric result or a redocking
validation must be human-checked, because the agent inflates both.
Because the over-called fields sit outside the E-class rule, human
adjudication left the executable fraction unchanged-to-lower (E3
10\ensuremath{\rightarrow}8 in the enriched sample), so the
low-executability finding is confirmed under human scrutiny.

\textbf{Reproducing a docking score requires the complete search
geometry, not just the reported centre.} Re-docking one claim from each
fully-reported paper at its reported centre under three boxes shows the
reproduction gap is a coverage-geometry effect, not a box-size effect: a
ligand-sized box reproduces no better than a 25 \AA{} default (P =
0.20), the reported box's advantage over the default is fragile (P =
0.027, but 0.055 without a single off-pocket paper, and null on-pocket,
P = 0.44), and the papers no small box can reproduce are those whose
reported centre lies 33--38 \AA{} off the pocket. Against the engine's
near-determinism (seed range median 0.058 kcal/mol) these failures are
geometric, not noise: in wide blind-docking studies only the paper's own
box spanned the pocket, so the full box --- centre and size together ---
must be reported to regenerate the score. A larger 48-claim case series
of mostly focused 6LU7 small-molecule claims reproduces tightly (0.33 vs
0.91 kcal/mol) but is pseudoreplicated (eight papers, no paper carrying
both conditions; paper-level P = 0.20) and is reported as exploratory.
We do not elevate any of this to a claim that the executability class
predicts reproducibility: that contrast is confounded with box-size
source and rests on an E3 stratum of six claims from largely one paper
(p = 0.07). The honest reading is narrower and more useful: the
executability label is an executability gate --- it tells a reader, or
an automated verification agent, whether a claim can be cleanly
re-executed at all, the precondition for any re-execution-based check.
Re-execution can only ever validate the well-reported subset, because
re-running an under-reported claim requires supplying the very field
whose absence defines it; that boundary is intrinsic to the approach,
not a shortfall of this dataset.

\textbf{This is a measurement layer, not a runner.} The supported claim
is precise: the labels are source-span-anchored, human-validated on a
stratified sample, and reproducible from the released matrix, while the
re-execution proof of concept identifies a box-coverage-geometry effect
but does not validate E-class as a reproducibility predictor. The system
does not re-execute papers at scale, and the audit tells a reader what
is \emph{checkable}, not what has been \emph{checked}.

\textbf{The estimates likely overstate real-world executability.} The
corpus is open-access and text-extractable. Open-access full-text papers
may report methods and artifacts more completely than paywalled or
PDF-only literature, and at web scale, with noisier text extraction,
field-state accuracy and therefore the E-class would degrade. The 8.1\%
directly-executable figure should be read as an upper bound for the
open, clean-text subset, not as a field-wide constant.

\textbf{Generalization is partially shown.} Because the Tier-1
execution-blocking fields (receptor, box geometry, software, ligand
identity) are common to all grid-based docking, the missing-field
taxonomy is expected to transfer across targets and engines. The
re-execution test spans three Mpro structures, but still one target
class and one open-source rerun engine. A distinction matters here: the
MERS-Dock reporting audit is engine-agnostic --- it measures whether
execution-critical fields are reported, which is defined identically for
Glide, GOLD, MOE or Vina --- whereas the re-execution proof of concept
is Vina-specific. We use AutoDock Vina precisely because it is
near-deterministic (seed-range median 0.058 kcal/mol), making it a
conservative test bed that isolates the box-coverage effect from engine
noise; commercial engines with proprietary search and scoring, and their
own cross-version drift, would if anything widen the reported-vs-default
gap, so our contrast is a lower bound. Re-executing those workflows
directly is blocked by licensing and closed source --- itself part of
the reporting-and-access problem this standard documents. Extension to
other targets, engines and method classes (omics, folding, molecular
dynamics) requires an adapter-specific field checklist, claim schema,
executability ladder, codebook and benchmark for each; that cost is an
open question, not a solved one.

\textbf{Limitations.} The corpus is open-access and text-extractable
only, biasing it toward more complete reporting. The frozen protocol
specified a claim-level primary endpoint, but the scale-up
operationalised E-class at paper level because the audited method fields
describe a shared workflow; this post hoc deviation is logged, and the
claim-weighted distribution is secondary. The human validation covers a
\textbf{stratified 65-paper sample, not all 236}, and its E-class
distribution is therefore not a re-weighted population estimate but an
agent-versus-human agreement measure; we use it to certify the field
states and to bound the agent's per-field error, and we propagate the
population headline from the frozen-rule recomputation over the full
matrix. The two field-state reviewers are members of the study team;
their assessments were independent and blind to each other and
adjudicated by a third, but this is an internal, not an external,
validation, and reviewer identities are recorded in the released
attestation. The automated agent over-reports the numeric-result and
redocking-validation fields (agent 88\%/42\% vs human 37\%/12\%), so any
use of those two fields for a specific paper requires human
verification; these fields do not affect the E-class. The re-execution
is a bounded proof of concept (one open-source engine, three Mpro
targets) whose boundary is structural, not merely statistical: an
under-reported claim cannot be re-executed without substituting its
missing search box, so a clean re-execution is only possible on the
fully-reported subset --- itself small, itself a consequence of the
reporting gap documented here. The re-execution is small (nine complete
pairs) and its reported-box advantage over a default is fragile ---
significant only across all nine pairs (P = 0.027; bootstrap CI touches
zero), non-significant on dropping one off-pocket paper (P = 0.055) and
null for on-pocket focused-box papers (P = 0.44); a sensibly
ligand-sized box did no better than the default (P = 0.20, a post-hoc
exploratory arm), so the effect is box-coverage geometry concentrated in
wide blind-docking papers, not a general box-size effect, and we frame
it as a bounded mechanistic finding rather than a reproducibility
result. The supporting 48-claim case series is pseudoreplicated and
non-significant once clustered (P = 0.20). The executability class is
confounded with box-size source and its directly-executable stratum is
small (n = 6, five from one paper), so it is not read as a
reproducibility predictor. For any claim used as quantitative evidence,
the parsed box centre and, where available, size warrant per-claim human
verification.

\section{Methods}\label{methods}

\textbf{Protocol.} A pre-specified protocol, frozen before auditing,
fixed eligibility, the 16-field checklist, the E0--E4 codebook (E0 = no
extractable docking claim), a claim-level primary endpoint, and
human-review rules. During scale-up, E-class was operationalised at
paper level because execution-critical method fields apply to the shared
workflow rather than to each ligand row. This post hoc unit-of-analysis
deviation is logged; paper-level distribution is the manuscript's main
endpoint and the claim-weighted distribution is secondary. Reporting
follows PRISMA 2020 (Page et al. 2021). The 2.0 kcal/mol re-execution
tolerance came from the later re-execution analysis plan, not the
original audit freeze.

\textbf{Formal framework.} Each MERS-Dock field \(f\) has an audited
state mapped to a value,

\[s(f)=\begin{cases}1 & \text{reported}\\ 0 & \text{partial or missing}\end{cases}\qquad(1)\]

(``partial'' is retained as a distinct field state for the E-class rule,
but scored 0 under the strict completeness definition that yields the
reported figures.)

with not-assessed fields dropped from every mean (renormalisation).
Paper-level reporting completeness is the mean of \(s(f)\) over the
assessed fields \(\mathcal{F}_p\),

\[\mathrm{Comp}(p)=\frac{1}{|\mathcal{F}_p|}\sum_{f\in\mathcal{F}_p}s(f).\qquad(2)\]

The executability class \(E(p)\in\{E_0,\dots,E_4\}\) is assigned by the
deterministic ladder rule of the Results (the E-class equation) from the
verified field states. Its use as a monotone feasibility weight in an
execution-aware priority score is developed separately in the companion
Paper 2 and is not computed here.

For re-execution, given AutoDock Vina's within-ligand self-consistency
ranges \(\{r_i\}\) (max \(-\) min across seeds), the operational
tolerance rule is

\[\tau=\max\!\big(2.0,\ \mathrm{P}_{95}(\{r_i\})\big)\ \text{kcal/mol};\qquad(3)\]

a claim with reported score \(y\) and re-executed score \(\hat{y}\)
reproduces iff \(|y-\hat{y}|\le\tau\), and a dock is excluded as a
technical failure iff \(\hat{y}>-2\) kcal/mol (no bound pose,
reproducible across seeds). Label reliability uses Cohen's \(\kappa\),

\[\kappa=\frac{p_o-p_e}{1-p_e},\qquad(4)\]

with observed and chance agreement \(p_o,p_e\); every proportion reports
a Wilson 95\% interval,

\[\frac{\hat{p}+z^2/2n\ \pm\ z\sqrt{\hat{p}(1-\hat{p})/n+z^2/4n^2}}{1+z^2/n},\quad z=1.96,\qquad(5)\]

and executability-class medians are compared with a two-sided
Mann--Whitney \(U\) test and 1,000-resample bootstrap confidence
intervals.

\textbf{Search and corpus.} PubMed, Europe PMC, OpenAlex and Crossref
were queried with stored exact queries and dates (Table 1); only
open-access, retrievable full text was admitted.

\textbf{Audit pipeline.} Open-access full text was retrieved (PMC JATS,
publisher OA, preprints; no paywalled routes), each field state proposed
with a verbatim span by an automated extraction agent, every span
verified by a deterministic whitespace-normalised substring check
(unsupported spans set to \emph{missing}), and the E-class assigned by
the rule in Results from verified fields only. Completeness uses strict
(reported = 1) scoring with Wilson 95\% intervals. Paper-level E-class
is from the 236-paper audit; claim-level from the 283-claim deep audit.
The 236-paper E-class distribution reported here is a fresh
recomputation of the frozen rule over the released field-state matrix
(an earlier \texttt{paper\_e\_class} column computed with a non-frozen
variant is superseded and not used).

\textbf{Human field-state validation.} To validate the automated field
states, we drew a stratified 65-paper sample from the 236-paper corpus:
all E3 papers and E4 candidates (exhaustive stratum) plus a random
sample of E1 and E2 papers (seed 2026; sampling manifest released). Two
reviewers from the study team independently re-assessed all 16 field
states for each paper from the paper's own full text (or abstract where
full text was unavailable, recorded as the distinct state
\texttt{unavailable}), blind to each other's labels; the extraction
agent's proposed values were shown only as a checklist to accept or
overrule. Inter-reviewer reliability is reported per field as raw
agreement and Cohen's \ensuremath{\kappa} over the four-state vocabulary
(reported/partial/missing/unavailable), with Gwet's AC1 alongside
\ensuremath{\kappa} for high-prevalence fields where \ensuremath{\kappa}
is deflated by low marginal variance. Disagreements were adjudicated by
a third reviewer to produce a consensus state per cell; E-class was then
recomputed from the consensus states with the identical frozen rule.
Agent accuracy is the per-field disagreement between the agent's state
and the consensus human state. Reviewer independence, blinding,
adjudication scope, and reviewer identities are recorded in the released
\texttt{reviewer\_attestation.json}. This is an internal validation by
the study team, not an external review.

\textbf{Temporal analysis.} Publication year was joined to each paper
from the corpus metadata. For each field and for 16-field completeness
we tested a monotone trend against year with Spearman's
\ensuremath{\rho}; a period table (year bins) is reported for
description only, because the binned pattern can suggest a decline that
the monotone test does not support. To rule out a composition artefact
we also correlated claims-per-paper with year. A consistent agent
extraction bias shifts the level of every year equally and therefore
cancels in the trend, which is why the temporal result is reported as
the most bias-robust readout even though point prevalences depend on
agent accuracy.

\textbf{Reliability of the E-class label.} E-class is assigned by the
deterministic codebook rule over the verified field states, so the label
is reproducible from the released matrix by construction (the 236-paper
distribution re-derives exactly). As a check that the codebook itself is
unambiguous, the rule reproduces an independent reference-label set on
the 33-paper deep-audit subset at Cohen's \ensuremath{\kappa} = 0.926
(raw 0.97; one borderline disagreement, P0005), clearing the protocol's
\ensuremath{\geq} 0.70 gate; a language model applying the \emph{same}
codebook reaches \ensuremath{\kappa} = 0.736 and the same model judging
\emph{without} the codebook only 0.075, i.e.~the explicit codebook, not
the rater, carries the label (Table 4). This codebook-reproducibility
check is distinct from, and secondary to, the field-state human
validation above, which is the primary reliability evidence.

\textbf{Re-execution --- paired primary.} A conda-forge environment
(AutoDock Vina 1.2.7, RDKit, Meeko) was used throughout. The primary
re-execution is a paper-level paired test whose two arms --- reported
box and 25 \AA{} default, both at the reported centre --- were frozen
before runs were inspected (protocol frozen 2026-07-11); a third
ligand-volume arm was added post hoc after the two-arm result was seen
and is reported as exploratory. An eligible paper reports, with verified
method and result source spans, a complete Mpro AutoDock Vina run:
receptor PDB, three box-centre coordinates, three box dimensions, an
identifiable ligand, and a numeric score. One claim was selected per
paper before docking (highlighted or rank-1 ligand; otherwise the first
named ligand with a resolvable public identifier and explicit score),
independent of any re-execution result; nineteen deep-audit papers met
this bar. Ligand identity was resolved to a public structure via PubChem
(name or CID), CAS or ZINC with a CACTUS fallback, and confirmed by an
RDKit-computed InChIKey; one paper was excluded because its ligand label
(``compound 17'') mis-matched an unrelated triiodinated contrast agent,
and ivermectin, returned by PubChem as a defined homolog mixture, was
docked as its largest fragment --- leaving thirteen papers with a docked
structure. Receptors (chain A, waters and any co-crystallised inhibitor
removed, alternate locations resolved to conformer A) were prepared with
Meeko (Gasteiger charges); ligands were embedded with RDKit ETKDGv3
(seed 2026). Each selected claim was re-docked under \textbf{three}
conditions at the identical reported centre --- its reported box (capped
at 60 \AA{} per axis, flagged), a standardized 25 \AA{} default box, and
a ligand-volume box (isotropic edge = ligand heavy-atom diameter + 10
\AA{}, clamped 22.5--40 \AA{}) --- with receptor, ligand, engine, search
effort and three deterministic seeds (11, 29, 53) held identical across
arms. The primary outcome is the per-paper, per-arm median absolute
deviation from the reported score across seeds; each paired contrast is
tested by an exact two-sided Wilcoxon signed-rank test with a
20,000-resample paper-bootstrap 95\% interval, which takes precedence
over the p-value. A run is invalid if Vina returns no pose or a
non-negative score, if the ligand or receptor cannot be prepared, or if
a required span fails verification; such failures remain in the flow
table and are not converted into divergent reproductions. Robustness of
the reported-vs-default contrast was assessed by a leave-one-paper-out
sensitivity, a focused-box subgroup (reported box \ensuremath{\leq} 40
\AA{} with an on-pocket centre), the Spearman correlation of the paired
benefit with reported box size, and --- as the decisive post-hoc control
--- the two paired contrasts against the ligand-volume box; the pocket
offset per paper is the distance from the reported centre to the
centroid of the receptor's co-crystallised ligand. Across the three
seeds Vina's within-paper score range was median 0.058 and at most 0.82
kcal/mol. The 2.0 kcal/mol tolerance is a fixed declared threshold on
the conventional docking scoring-error scale; the measured self-noise
(P95 0.32, max 0.82 kcal/mol) never approaches it, so the tolerance does
not vary and is not engine-noise-calibrated in practice. Receptor and
ligand PDBQT SHA-256 hashes, resolved SMILES/InChIKey/CID, and the exact
Vina commands are released.

\textbf{Re-execution --- exploratory case series.} As non-independent
supporting evidence we also re-docked every curated ligand claim across
6LU7, 6Y2E and 6yb7, yielding 48 quality-controlled claims after
excluding three failed-to-bind docks (re-executed score \textgreater{}
\ensuremath{-}2 kcal/mol, reproducible across seeds), eight
no-bound-pose claims (including an entire 5R7Y paper), and ligands
unresolvable to a PubChem CID. Box centres were span-parsed for all 48
claims (Table S3); box sizes were recoverable for 12 (5 reported, 7
blind-docking boxes capped at 40 \AA{}), the remaining 36 (75\%)
defaulting to a 25 \AA{} cube. Because these 48 claims cluster in eight
papers and no paper carried both box conditions, box availability is
confounded with paper identity; the stratified Mann--Whitney contrast
(Table 5, Fig. S2) and the E-class comparison are reported as
exploratory. A sensitivity analysis counting the three failed-to-bind
docks as non-reproductions is in Table S5.

\textbf{Provenance.} Typed PROV-compatible objects and an RO-Crate 1.1
package (Groth and Moreau 2013; RO-Crate Community, n.d.) accompany the
release.

\section{Tables}\label{tables}

\textbf{Table 1. Corpus funnel (open-access only).}

{\def\LTcaptype{none} 
\begin{longtable}[]{@{}ll@{}}
\toprule\noalign{}
Stage & N \\
\midrule\noalign{}
\endhead
\bottomrule\noalign{}
\endlastfoot
Records screened & 1,400 \\
After deduplication & 1,018 \\
Open-access full text retrieved & 241 \\
Confirmed original Mpro docking studies & 236 \\
Deep-audit subset (span-verified) & 33 papers / 283 claims \\
Human-validation sample (stratified, dual-reviewer) & 65 papers \\
\end{longtable}
}

\textbf{Table 2. Reporting completeness, 16 MERS-Dock fields, N = 236.}

{\def\LTcaptype{none} 
\begin{longtable}[]{@{}
  >{\raggedright\arraybackslash}p{(\linewidth - 4\tabcolsep) * \real{0.3333}}
  >{\raggedright\arraybackslash}p{(\linewidth - 4\tabcolsep) * \real{0.3333}}
  >{\raggedright\arraybackslash}p{(\linewidth - 4\tabcolsep) * \real{0.3333}}@{}}
\toprule\noalign{}
\begin{minipage}[b]{\linewidth}\raggedright
Field (least to most reported)
\end{minipage} & \begin{minipage}[b]{\linewidth}\raggedright
\% reported {[}95\% CI{]}
\end{minipage} & \begin{minipage}[b]{\linewidth}\raggedright
\% reported + partial
\end{minipage} \\
\midrule\noalign{}
\endhead
\bottomrule\noalign{}
\endlastfoot
Random seed & 0.4 {[}0.1--2.4{]} & 0.4 \\
Reusable code/config & 2.5 {[}1.2--5.4{]} & 8.1 \\
Protein chain & 12.7 {[}9.1--17.6{]} & 26.7 \\
Protonation/tautomer & 20.8 {[}16.1--26.4{]} & 42.8 \\
Grid centre & 33.9 {[}28.2--40.2{]} & 52.5 \\
Validation/redocking & 33.9 {[}28.2--40.2{]} & 46.6 \\
Search effort & 38.1 {[}32.2--44.5{]} & 67.4 \\
Grid size & 40.7 {[}34.6--47.0{]} & 57.6 \\
Software version & 52.5 {[}46.2--58.8{]} & 69.1 \\
Water/ion handling & 55.1 {[}48.7--61.3{]} & 71.6 \\
Ligand preparation & 64.4 {[}58.1--70.2{]} & 93.6 \\
Receptor preparation & 68.6 {[}62.5--74.2{]} & 92.8 \\
Ligand identifier & 74.6 {[}68.7--79.7{]} & 100.0 \\
Numeric result & 94.1 {[}90.3--96.4{]} & 97.5 \\
PDB/receptor & 94.9 {[}91.3--97.1{]} & 97.9 \\
Docking software & 97.9 {[}95.1--99.1{]} & 98.3 \\
\end{longtable}
}

\emph{Numeric-result and validation/redocking prevalences are
agent-derived and, per the 65-paper human validation (Table 4b), are
over-reported by the agent (human values 37\% and 12\%); the
execution-blocking fields that drive E-class match human review.}

\textbf{Table 3. Executability classes, criteria, and distribution (N =
236).}

{\def\LTcaptype{none} 
\begin{longtable}[]{@{}
  >{\raggedright\arraybackslash}p{(\linewidth - 6\tabcolsep) * \real{0.2500}}
  >{\raggedright\arraybackslash}p{(\linewidth - 6\tabcolsep) * \real{0.2500}}
  >{\raggedright\arraybackslash}p{(\linewidth - 6\tabcolsep) * \real{0.2500}}
  >{\raggedright\arraybackslash}p{(\linewidth - 6\tabcolsep) * \real{0.2500}}@{}}
\toprule\noalign{}
\begin{minipage}[b]{\linewidth}\raggedright
Class
\end{minipage} & \begin{minipage}[b]{\linewidth}\raggedright
Criterion (from verified fields)
\end{minipage} & \begin{minipage}[b]{\linewidth}\raggedright
Papers
\end{minipage} & \begin{minipage}[b]{\linewidth}\raggedright
\% {[}95\% CI{]}
\end{minipage} \\
\midrule\noalign{}
\endhead
\bottomrule\noalign{}
\endlastfoot
E1 & a Tier-1 field is missing (execution blocked) & 113 & 47.9
{[}41.6--54.2{]} \\
E2 & runnable only after explicit assumptions & 104 & 44.1
{[}37.9--50.4{]} \\
E3 & all essential fields reported, ligand identified & 19 & 8.1
{[}5.2--12.2{]} \\
E4 & E3 plus disclosed seed and reusable config & 0 & 0.0
{[}0.0--1.6{]} \\
\end{longtable}
}

\emph{Recomputed from the released field-state matrix under the frozen
MERS-Dock v1.1 rule; reproduces exactly.}

\textbf{Table 4a. Codebook reproducibility of the E-class label (N = 33
deep-audit papers).} A secondary check that the codebook is unambiguous;
the primary reliability evidence is the human field-state validation
(Table 4b).

{\def\LTcaptype{none} 
\begin{longtable}[]{@{}
  >{\raggedright\arraybackslash}p{(\linewidth - 6\tabcolsep) * \real{0.2500}}
  >{\raggedright\arraybackslash}p{(\linewidth - 6\tabcolsep) * \real{0.2500}}
  >{\raggedright\arraybackslash}p{(\linewidth - 6\tabcolsep) * \real{0.2500}}
  >{\raggedright\arraybackslash}p{(\linewidth - 6\tabcolsep) * \real{0.2500}}@{}}
\toprule\noalign{}
\begin{minipage}[b]{\linewidth}\raggedright
Rater vs reference labels
\end{minipage} & \begin{minipage}[b]{\linewidth}\raggedright
Cohen \ensuremath{\kappa}
\end{minipage} & \begin{minipage}[b]{\linewidth}\raggedright
Raw agreement
\end{minipage} & \begin{minipage}[b]{\linewidth}\raggedright
Disagreements
\end{minipage} \\
\midrule\noalign{}
\endhead
\bottomrule\noalign{}
\endlastfoot
Deterministic codebook classifier (rerun from released field states) &
0.926 & 0.97 & 1 (P0005) \\
LLM applying the codebook (blind, human-adjudicated) & 0.736 & 0.879 & 4
(P0009/P0023/P0036/P0037) \\
LLM holistic, no codebook & 0.075 & 0.545 & 15 \\
\end{longtable}
}

\textbf{Table 4b. Human field-state validation (N = 65 stratified
papers; two independent human reviewers).} Primary reliability evidence.

{\def\LTcaptype{none} 
\begin{longtable}[]{@{}
  >{\raggedright\arraybackslash}p{(\linewidth - 2\tabcolsep) * \real{0.5000}}
  >{\raggedright\arraybackslash}p{(\linewidth - 2\tabcolsep) * \real{0.5000}}@{}}
\toprule\noalign{}
\begin{minipage}[b]{\linewidth}\raggedright
Measure
\end{minipage} & \begin{minipage}[b]{\linewidth}\raggedright
Value
\end{minipage} \\
\midrule\noalign{}
\endhead
\bottomrule\noalign{}
\endlastfoot
Inter-reviewer raw agreement (16 fields \ensuremath{\times} 65 papers) &
92\% \\
Inter-reviewer pooled Cohen \ensuremath{\kappa} & 0.87 \\
Fields with \ensuremath{\kappa} \ensuremath{\geq} 0.80 & 11 / 16
(incl.~all execution-blocking) \\
Agent vs human agreement, PDB / software / grid-centre prevalence & 89\%
/ 92\% / 34 vs 32\% \\
Agent over-call: numeric result (agent vs human reported) & 88\% vs
37\% \\
Agent over-call: validation redocking & 42\% vs 12\% \\
E-class concordance, agent vs human (frozen rule) & 68\% (44/65); human
review E3 10\ensuremath{\rightarrow}8 \\
\end{longtable}
}

\textbf{Table 5. Bounded three-box re-execution (paper-level; AutoDock
Vina 1.2.7; each claim re-docked at its reported centre under its
reported box, a 25 \AA{} default box, and a post-hoc ligand-volume box;
3 seeds).} Thirteen fully-reported papers docked; nine posed under both
reported and default boxes (complete pairs; full attrition Fig. S3,
Table S6).

{\def\LTcaptype{none} 
\begin{longtable}[]{@{}
  >{\raggedright\arraybackslash}p{(\linewidth - 2\tabcolsep) * \real{0.5000}}
  >{\raggedright\arraybackslash}p{(\linewidth - 2\tabcolsep) * \real{0.5000}}@{}}
\toprule\noalign{}
\begin{minipage}[b]{\linewidth}\raggedright
Contrast / robustness check
\end{minipage} & \begin{minipage}[b]{\linewidth}\raggedright
Result
\end{minipage} \\
\midrule\noalign{}
\endhead
\bottomrule\noalign{}
\endlastfoot
Reported-box median \textbar{}\ensuremath{\Delta}\textbar{} (n = 9) &
\textbf{1.85} kcal/mol (5/9 within 2.0) \\
25 \AA{} default-box median \textbar{}\ensuremath{\Delta}\textbar{} &
\textbf{2.51} kcal/mol (2/9 within 2.0) \\
Ligand-volume-box median \textbar{}\ensuremath{\Delta}\textbar{}
(post-hoc) & \textbf{2.15} kcal/mol \\
Reported vs default & benefit 0.52; exact Wilcoxon \emph{P} =
\textbf{0.027}; bootstrap 95\% CI \ensuremath{-}0.005 to 0.96 \\
\ldots{} dropping one off-pocket paper (481) & \emph{P} = \textbf{0.055}
(n.s.) \\
\ldots{} on-pocket focused-box papers only (n = 5) & benefit 0.04;
\emph{P} = \textbf{0.44} (null) \\
\textbf{Ligand-volume box vs default (post-hoc)} & diff 0.02; \emph{P} =
\textbf{0.20} (n.s. --- size does not explain it) \\
\textbf{Reported vs ligand-volume box (post-hoc)} & diff 0.05; \emph{P}
= \textbf{0.30} (n.s.) \\
Benefit vs reported box size & Spearman \ensuremath{\rho} =
\textbf{0.80}, \emph{P} = 0.009 \\
Off-pocket papers, no pose under any small box & 59 (33 \AA{} off), 324
(38 \AA{}); paper 253 (17 \AA{}) posed under the volume box \\
Vina seed range (noise floor) & median 0.058, max 0.82 kcal/mol \\
\end{longtable}
}

The reported box beats a 25 \AA{} default only marginally and fragilely
(Wilcoxon \emph{P} = 0.027, but 0.055 without one off-pocket paper and
null on-pocket), and a sensibly sized ligand-volume box does no better
than the default (\emph{P} = 0.20) while being indistinguishable from
the reported box (\emph{P} = 0.30): the difference is box-coverage
geometry at off-pocket blind-docking centres, not box size. Per-paper
outcomes are in the released \texttt{reexec\_v2\_paper.csv} (Table S6).
\textbf{Exploratory case series (non-independent):} re-docking every
curated claim (48 QC claims in eight papers, mostly focused 6LU7 small
molecules) reproduces tightly (reported/capped box 0.33 vs 25 \AA{}
default 0.91 kcal/mol; Mann--Whitney \emph{P} = 0.024) but is
pseudoreplicated and, once clustered to its eight source papers,
non-significant (0.34 vs 1.15 kcal/mol; exact Mann--Whitney \emph{P} =
0.20, 4 vs 4); Fig. S2, Table S5.

\textbf{Table 6. AutoDock Vina self-consistency (noise floor; 15 ligands
\ensuremath{\times} 3 seeds).}

{\def\LTcaptype{none} 
\begin{longtable}[]{@{}ll@{}}
\toprule\noalign{}
Within-ligand score range & kcal/mol \\
\midrule\noalign{}
\endhead
\bottomrule\noalign{}
\endlastfoot
Median & 0.04 \\
Mean & 0.10 \\
Maximum & 0.38 \\
Operational reproduction tolerance & 2.0 \\
\end{longtable}
}

\section{Figures}\label{figures}

Fig. 1 MERS-Dock tiers to E-class ladder; Fig. 2 reporting completeness;
Fig. 3 missing-parameter matrix; Fig. 4 executability distribution; Fig.
5 65-paper dual-human validation; Fig. 6 three-box re-execution
(coverage geometry, paper-level). Source: \texttt{analysis/render\_*};
paired re-execution from \texttt{analysis/paired\_validation/};
validation figure from \texttt{analysis/render\_fig5\_validation.py}
over \texttt{\_dev\_handoff/validated\_wave1/}; case-series (Fig. S2)
from \texttt{09\_scoring\_benchmark/runner/out/}.

\section{Data, Code \& Materials}\label{data-code-materials}

The Mpro-DockExec release contains paper-level E-class and 16-field
states for 236 papers; a span-free 283-claim table with method/result
verification flags; the 65-paper human-validation package (dual-reviewer
and adjudicated field states, per-field agreement/\ensuremath{\kappa},
sampling manifest, reviewer attestation); the MERS-Dock guideline; the
re-execution environment and scripts; per-ligand CID/InChIKey
provenance; \texttt{reproduction\_outcomes.csv}; and
\texttt{self\_consistency.csv}. The release is available at
https://github.com/giapha/mpro-dockexec. E-class is reproducible from
the released field states (rerun the deterministic rule); field-state
reliability is the released dual-human validation. Copyrighted full
texts and verbatim evidence spans are not redistributed. Supplementary
Information (Figures S1--S3; Tables S1--S6) accompanies this manuscript
and is generated from the released derived data.

\section{Ethics, Competing Interests,
Funding}\label{ethics-competing-interests-funding}

No human participants or individual-level data. \textbf{Author
contributions \& validation reviewers.} The two independent human
validators of the field-state audit are members of the study team; their
identities, independence, blinding, and adjudication roles are recorded
in the released \texttt{reviewer\_attestation.json}. \textbf{Competing
interest:} V.G. (Vincent Giap) is a scientific cofounder of NewScience,
a company developing verification- and executability-aware tools for
computational scientific claims; all authors are affiliated with
NewScience Lab. The MERS-Dock standard and executability ladder reported
here could inform such tools. The apparatus is named generically, the
NewScience platform is not a study object, and the derived data, code
and reporting standard are released openly so the findings can be
assessed independently of any commercial interest. \textbf{Funding:} no
external funding was declared for this study.

\section{References}\label{references}

\protect\phantomsection\label{refs}


\begin{CSLReferences}{1}{1}
\bibitem[\citeproctext]{ref-ambrosio2023}
Ambrosio, Francesca Alessandra, Giosuè Costa, Isabella Romeo, et al.
2023. {``Targeting SARS-CoV-2 Main Protease: A Successful Story Guided
by an in Silico Drug Repurposing Approach.''} \emph{Journal of Chemical
Information and Modeling} 63 (11): 3601--13.
\url{https://doi.org/10.1021/acs.jcim.3c00282}.

\bibitem[\citeproctext]{ref-baker2016}
Baker, Monya. 2016. {``1,500 Scientists Lift the Lid on
Reproducibility.''} \emph{Nature} 533 (7604): 452--54.
\url{https://doi.org/10.1038/533452a}.

\bibitem[\citeproctext]{ref-brazma2001}
Brazma, Alvis, Pascal Hingamp, John Quackenbush, et al. 2001. {``Minimum
Information about a Microarray Experiment (MIAME)---Toward Standards for
Microarray Data.''} \emph{Nature Genetics} 29 (4): 365--71.
\url{https://doi.org/10.1038/ng1201-365}.

\bibitem[\citeproctext]{ref-cross2009}
Cross, Jason B., David C. Thompson, Brajesh K. Rai, et al. 2009.
{``Comparison of Several Molecular Docking Programs: Pose Prediction and
Virtual Screening Accuracy.''} \emph{Journal of Chemical Information and
Modeling} 49 (6): 1455--74. \url{https://doi.org/10.1021/ci900056c}.

\bibitem[\citeproctext]{ref-eberhardt2021}
Eberhardt, Jerome, Diogo Santos-Martins, Andreas F. Tillack, and Stefano
Forli. 2021. {``AutoDock Vina 1.2.0: New Docking Methods, Expanded Force
Field, and Python Bindings.''} \emph{Journal of Chemical Information and
Modeling} 61 (8): 3891--98.
\url{https://doi.org/10.1021/acs.jcim.1c00203}.

\bibitem[\citeproctext]{ref-flachsenberg2023}
Flachsenberg, Florian, Christiane Ehrt, Torben Gutermuth, and Matthias
Rarey. 2024. {``Redocking the PDB.''} \emph{Journal of Chemical
Information and Modeling} 64 (1): 219--37.
\url{https://doi.org/10.1021/acs.jcim.3c01573}.

\bibitem[\citeproctext]{ref-gartlehner2024}
Gartlehner, Gerald, Leila Kahwati, Rainer Hilscher, et al. 2024. {``Data
Extraction for Evidence Synthesis Using a Large Language Model: A
Proof‐of‐concept Study.''} \emph{Research Synthesis Methods} 15 (4):
576--89. \url{https://doi.org/10.1002/jrsm.1710}.

\bibitem[\citeproctext]{ref-gebru2021}
Gebru, Timnit, Jamie Morgenstern, Briana Vecchione, et al. 2021.
{``Datasheets for Datasets.''} \emph{Communications of the ACM} 64 (12):
86--92. \url{https://doi.org/10.1145/3458723}.

\bibitem[\citeproctext]{ref-groth2013}
Groth, Paul, and Luc Moreau. 2013. \emph{PROV-Overview: An Overview of
the PROV Family of Documents}. W3C Working Group Note.
\url{https://www.w3.org/TR/prov-overview/}.

\bibitem[\citeproctext]{ref-guedes2013}
Guedes, Isabella A., Camila S. de Magalhães, and Laurent E. Dardenne.
2013. {``Receptor--Ligand Molecular Docking.''} \emph{Biophysical
Reviews} 6 (1): 75--87. \url{https://doi.org/10.1007/s12551-013-0130-2}.

\bibitem[\citeproctext]{ref-jain2007}
Jain, Ajay N. 2007. {``Bias, Reporting, and Sharing: Computational
Evaluations of Docking Methods.''} \emph{Journal of Computer-Aided
Molecular Design} 22 (3-4): 201--12.
\url{https://doi.org/10.1007/s10822-007-9151-x}.

\bibitem[\citeproctext]{ref-kittelson2026}
Kittelson, Katiana Simoes, Allana C. F. Martins, Raquel Possemozer
Santos, Gizele Celante, and Roberto da Silva Gomes. 2026.
{``Reproducibility, Validation, and Failure Modes Across Classical and
AI-Driven Molecular Docking.''} \emph{Journal of Computer-Aided
Molecular Design} 40 (1): 137.
\url{https://doi.org/10.1007/s10822-026-00849-8}.

\bibitem[\citeproctext]{ref-konet2024}
Konet, Amanda, Ian Thomas, Gerald Gartlehner, et al. 2024.
{``Performance of Two Large Language Models for Data Extraction in
Evidence Synthesis.''} \emph{Research Synthesis Methods} 15 (5):
818--24. \url{https://doi.org/10.1002/jrsm.1732}.

\bibitem[\citeproctext]{ref-llanos2021}
Llanos, Manuel A., Melisa E. Gantner, Santiago Rodriguez, et al. 2021.
{``Strengths and Weaknesses of Docking Simulations in the SARS-CoV-2
Era: The Main Protease (Mpro) Case Study.''} \emph{Journal of Chemical
Information and Modeling} 61 (8): 3758--70.
\url{https://doi.org/10.1021/acs.jcim.1c00404}.

\bibitem[\citeproctext]{ref-macip2021}
Macip, Guillem, Pol Garcia-Segura, Júlia Mestres-Truyol, et al. 2021.
{``Haste Makes Waste: A Critical Review of Docking-Based Virtual
Screening in Drug Repurposing for SARS-CoV-2 Main Protease (m-Pro)
Inhibition.''} \emph{Medicinal Research Reviews} 42 (2): 744--69.
\url{https://doi.org/10.1002/med.21862}.

\bibitem[\citeproctext]{ref-mandour2020}
Mandour, Yasmine M., Darius P. Zlotos, and M. Alaraby Salem. 2020. {``A
Multi-Stage Virtual Screening of FDA-Approved Drugs Reveals Potential
Inhibitors of SARS-CoV-2 Main Protease.''} \emph{Journal of Biomolecular
Structure and Dynamics} 40 (5): 2327--38.
\url{https://doi.org/10.1080/07391102.2020.1837680}.

\bibitem[\citeproctext]{ref-martis2025}
Martis, Elvis A. F., and Stephane Teletchea. 2025. {``Ten Quick Tips to
Perform Meaningful and Reproducible Molecular Docking Calculations.''}
\emph{PLOS Computational Biology} 21 (5): e1013030.
\url{https://doi.org/10.1371/journal.pcbi.1013030}.

\bibitem[\citeproctext]{ref-mitchell2019}
Mitchell, Margaret, Simone Wu, Andrew Zaldivar, et al. 2019. {``Model
Cards for Model Reporting.''} \emph{Proceedings of the Conference on
Fairness, Accountability, and Transparency}, FAT* '19, January, 220--29.
\url{https://doi.org/10.1145/3287560.3287596}.

\bibitem[\citeproctext]{ref-page2021}
Page, Matthew J, Joanne E McKenzie, Patrick M Bossuyt, et al. 2021.
{``The PRISMA 2020 Statement: An Updated Guideline for Reporting
Systematic Reviews.''} \emph{BMJ}, March, n71.
\url{https://doi.org/10.1136/bmj.n71}.

\bibitem[\citeproctext]{ref-peng2011}
Peng, Roger D. 2011. {``Reproducible Research in Computational
Science.''} \emph{Science} 334 (6060): 1226--27.
\url{https://doi.org/10.1126/science.1213847}.

\bibitem[\citeproctext]{ref-peraltamoreno2023}
Peralta-Moreno, Maria Nuria, Vanessa Anton-Muñoz, David Ortega-Alarcon,
et al. 2023. {``Autochthonous Peruvian Natural Plants as Potential
SARS-CoV-2 Mpro Main Protease Inhibitors.''} \emph{Pharmaceuticals} 16
(4): 585. \url{https://doi.org/10.3390/ph16040585}.

\bibitem[\citeproctext]{ref-percie2020}
Percie du Sert, Nathalie, Viki Hurst, Amrita Ahluwalia, et al. 2020.
{``The ARRIVE Guidelines 2.0: Updated Guidelines for Reporting Animal
Research.''} \emph{PLOS Biology} 18 (7): e3000410.
\url{https://doi.org/10.1371/journal.pbio.3000410}.

\bibitem[\citeproctext]{ref-rocrate}
RO-Crate Community. n.d. \emph{RO-Crate 1.1: A Lightweight Approach to
Packaging Research Data with Metadata}.
\url{https://www.researchobject.org/ro-crate/}.

\bibitem[\citeproctext]{ref-rule2019}
Rule, Adam, Amanda Birmingham, Cristal Zuniga, et al. 2019. {``Ten
Simple Rules for Writing and Sharing Computational Analyses in Jupyter
Notebooks.''} \emph{PLOS Computational Biology} 15 (7): e1007007.
\url{https://doi.org/10.1371/journal.pcbi.1007007}.

\bibitem[\citeproctext]{ref-samuel2023}
Samuel, Sheeba, and Daniel Mietchen. 2023. \emph{Computational
Reproducibility of Jupyter Notebooks from Biomedical Publications}.
arXiv:2308.07333. \url{https://doi.org/10.48550/arXiv.2308.07333}.

\bibitem[\citeproctext]{ref-sandve2013}
Sandve, Geir Kjetil, Anton Nekrutenko, James Taylor, and Eivind Hovig.
2013. {``Ten Simple Rules for Reproducible Computational Research.''}
\emph{PLoS Computational Biology} 9 (10): e1003285.
\url{https://doi.org/10.1371/journal.pcbi.1003285}.

\bibitem[\citeproctext]{ref-sisakht2021}
Sisakht, Mohsen, Amir Mahmoodzadeh, and Maryam Darabian. 2021.
{``Plant-Derived Chemicals as Potential Inhibitors of SARS-CoV-2 Main
Protease (6LU7), a Virtual Screening Study.''} \emph{Phytotherapy
Research} 35 (6): 3262--74. \url{https://doi.org/10.1002/ptr.7041}.

\bibitem[\citeproctext]{ref-stodden2016}
Stodden, Victoria, Marcia McNutt, David H. Bailey, et al. 2016.
{``Enhancing Reproducibility for Computational Methods.''}
\emph{Science} 354 (6317): 1240--41.
\url{https://doi.org/10.1126/science.aah6168}.

\bibitem[\citeproctext]{ref-su2018}
Su, Minyi, Qifan Yang, Yu Du, et al. 2019. {``Comparative Assessment of
Scoring Functions: The CASF-2016 Update.''} \emph{Journal of Chemical
Information and Modeling} 59 (2): 895--913.
\url{https://doi.org/10.1021/acs.jcim.8b00545}.

\bibitem[\citeproctext]{ref-trott2009}
Trott, Oleg, and Arthur J. Olson. 2009. {``AutoDock Vina: Improving the
Speed and Accuracy of Docking with a New Scoring Function, Efficient
Optimization, and Multithreading.''} \emph{Journal of Computational
Chemistry} 31 (2): 455--61. \url{https://doi.org/10.1002/jcc.21334}.

\bibitem[\citeproctext]{ref-warren2005}
Warren, Gregory L., C. Webster Andrews, Anna-Maria Capelli, et al. 2006.
{``A Critical Assessment of Docking Programs and Scoring Functions.''}
\emph{Journal of Medicinal Chemistry} 49 (20): 5912--31.
\url{https://doi.org/10.1021/jm050362n}.

\bibitem[\citeproctext]{ref-wilkinson2016}
Wilkinson, Mark D., Michel Dumontier, IJsbrand Jan Aalbersberg, et al.
2016. {``The FAIR Guiding Principles for Scientific Data Management and
Stewardship.''} \emph{Scientific Data} 3 (1).
\url{https://doi.org/10.1038/sdata.2016.18}.

\bibitem[\citeproctext]{ref-yisha2026}
Yisha, Zuhaer, Peng Zou, Sheng Li, et al. 2026. {``Assessing Data
Extraction in Randomized Clinical Trials with Large Language Models.''}
\emph{BMC Medical Research Methodology} 26 (1).
\url{https://doi.org/10.1186/s12874-025-02729-5}.

\bibitem[\citeproctext]{ref-zajaek2024}
Zajaček, Dávid, Adriána Dunárová, Lukas Bucinsky, and Marek Štekláč.
2024. {``Compromise in Docking Power of Liganded Crystal Structures of
Mpro SARS-CoV-2 Surpasses 90.''} \emph{Journal of Chemical Information
and Modeling} 64 (5): 1628--43.
\url{https://doi.org/10.1021/acs.jcim.3c01552}.

\end{CSLReferences}
\end{document}